   \documentclass{aa}
   \usepackage{graphics}

\begin{document}

   \voffset 1.5 true cm

   \thesaurus{10     
              (08.03.3; 
               08.12.1;  
               10.05.1;  
               10.19.1)}

   \title{Chemical enrichment and star formation in the Milky Way disk}
   \subtitle{I. Sample description and chromospheric age--metallicity relation}

   \author{H. J. Rocha-Pinto\inst{1} 
          \and W. J. Maciel\inst{1}
          \and John Scalo\inst{2}
          \and Chris Flynn\inst{3}}

   \offprints{H. J. Rocha-Pinto}

   \institute{Instituto Astron\^omico e Geof\'{\i}sico, Universidade de S\~ao 
              Paulo, Av. Miguel Stefano 4200, 04301-904 S\~ao Paulo SP, 
              Brazil\\ emails: helio@iagusp.usp.br; maciel@iagusp.usp.br 
             \and Department of Astronomy, The University of Texas at Austin, 
              USA. email: parrot@astro.as.utexas.edu
             \and Tuorla Observatory, V\"ais\"al\"antie 20, FIN-21500, Piikki\"o, Finland. 
              email: cflynn@astro.utu.fi}
   
   \titlerunning{Chemical enrichment and formation of the Milky Way disk}
   \authorrunning{H.J. Rocha-Pinto et al.}

   \date{Received date; accepted date}
   
   \maketitle

   \begin{abstract}

   The age--metallicity relation of the solar neighbourhood is studied 
   using a sample of 552 late-type dwarfs. This sample was built from the intersection 
   of photometric catalogues with chromospheric activity surveys of the Mount Wilson 
   group. For these stars, metallicities were estimated from {\it uvby} data, and ages 
   were calculated from their chromospheric emission levels using a new 
   metallicity-dependent 
   chromospheric activity--age relation developed by Rocha-Pinto \& Maciel (\cite{RPM98}). 
   A careful estimate of the errors in the chromospheric age is made. The errors 
   in the chromospheric indices are shown to include partially the effects of the 
   stellar magnetic cycles, although a detailed treatment of this error is still beyond 
   our knowledge. It is shown that the results are not affected by the presence of 
   unresolved binaries in the sample. 
   We derive an age--metallicity relation which 
   confirms the mean trend found by previous workers. The mean metallicity shows a  
   slow, steady increase with time, amounting at least 0.56 dex in 15 Gyr. The initial 
   metallicity of the disk is around $-0.70$ dex, in agreement with the G dwarf 
   metallicity distribution (Rocha-Pinto \& Maciel \cite{RPM96}).
   According to our data, {\it the intrinsic cosmic dispersion in metal abundances 
   is around 0.13 dex}, a factor of two smaller than that found by Edvardsson et al. 
   (\cite{Edv}). We show that chromospheric ages are compatible with isochrone ages, within the 
   expected errors, so that the difference in the scatter cannot be caused by the accuracy 
   of our ages and metallicities. This reinforces 
   some suggestions that the Edvarsson et al.'s sample is not suitable to the 
   determination of the age--metallicity relation.

   \end{abstract}

\section{Introduction}

   In the general picture for the evolution of our Galaxy, the stars form 
   from gas enriched by previous stellar generations, and eject new 
   metals to the interstellar medium after their death. Due to this 
   continuing enrichment, old stars are likely to be metal-poorer than 
   younger stars.

   The age--metallicity relation (hereafter AMR) has been the subject of 
   many studies in 
   the literature. The first systematic attempt to determine this relation 
   was made by Twarog (\cite{twar}). He presented photometric metallicities 
   and isochrone ages for 329 F dwarfs, finding a smooth increasing relation 
   with an average scatter of 0.12 dex. Twarog's sample was reanalysed 
   subsequently by two different groups. Carlberg et al. (\cite{carl}) have 
   found a very flat AMR, probably because they have cut from the sample 
   all stars with metallicities lower than $-0.50$ dex. On the other hand, 
   Meusinger et al. (\cite{meu}) used updated isochrones and a metallicity 
   calibration and found an AMR very similar to that of Twarog. Other 
   attempts to derive this relation using photometric metallicities were made by 
   Ann \& Kang (\cite{annkang}) and Marsakov et al. 
   (\cite{marsakov}), but these works suffer from the lack of an unbiased  
   sample selection. 

   By far the most common approach to study the AMR is the use of photometric 
   metallicities and isochrone ages, since this allows for a large sample which 
   can compensate for a poor accuracy in 
   these quantities. However, some studies make use of spectroscopic 
   metallicities. Nissen et al. (\cite{nissen}) have presented [Fe/H] and ages for 29 
   F dwarfs taken from the larger sample that was investigated later in more detail by 
   Edvardsson et al. (\cite{Edv}, hereafter Edv93). Their data agrees well with 
   Twarog's AMR, although the scatter seems to be higher. Lee et al. (\cite{lee}) give 
   a spectroscopic AMR for 559 disk stars whose metallicities are given in the 
   catalogue of Cayrel de Strobel et al. (\cite{cayrel85}). The authors measure ages 
   from isochrones in five different diagrams, some of which are likely to be 
   independent of the stellar distance, which is one of the major sources of error in 
   the isochrone age determination (see for example Ng \& Bertelli \cite{ng}). 
   However, their sample is highly heterogeneous, not only regarding the 
   metallicity sources, but also the spectral types used for the study.

   Presently, the most significant work on the AMR was done by Edv93. They measured 
   accurate spectroscopic metallicities on 189 carefully selected disk stars. Ages 
   were found by VandenBerg's (\cite{VandenBerg}) isochrones. Their result is rather 
   surprising: while the mean AMR is similar to Twarog's AMR, the 
   metallicity dispersion is so high that it casts doubts about the real meaning of 
   the age--metallicity relation. The same data were recently reanalysed by Ng \& 
   Bertelli (\cite{ng}), using updated isochrones as well as HIPPARCOS parallaxes, 
   and the high dispersion in metallicity was confirmed. 
   A high metallicity dispersion can also be found in the AMR of open clusters 
   (Strobel \cite{strobel}; Janes \& Friel \cite{friel}; 
   Carraro \& Chiosi \cite{carchi}; Carraro et al. 
   \cite{carraro}). 

   Although this seemed puzzling with respect to the previous 
   well-marked photometric AMR, it has given rise to a new era in studies about the 
   chemical evolution of the disk in which the gas is not very well mixed so that 
   local inhomogeneities would obscure the overall growth of the metallicity.  

   Nevertheless, the AMR is still a poorly known function. A critical review of 
   the literature shows that only two independent carefully selected samples of 
   field disk stars were 
   ever used in its study: Twarog's and Edv93's samples. And the conclusions 
   that can be drawn from both samples, mainly on the metallicity dispersion, are 
   in remarkable disagreement, strongly motivating the present study. 
   One other aspect that deserves further investigation is the apparent 
   lack of metal rich objects in 
   the age range 3--5 Gyr as pointed out by Carraro et al. (\cite{carraro}), since 
   a similar feature is also marginally visible in the data of Carlberg et al. 
   (\cite{carl}). 

   This work is the first of a series in which we revisit some fundamental 
   constraints on chemical evolution, by using an independent and extensive 
   sample of long-lived dwarfs. The main novelty of these papers is an attempt 
   to explore an independent tool to measure 
   stellar ages: the chromospheric activity level, which can be even more accurate to 
   measure ages of late-type stars than isochrones (Lachaume et al. \cite{lach}). 
   The only study ever published using this approach is that 
   of Barry (\cite{barry}) but the sample selection prevents his AMR to be taken 
   as representative. There are several chromospheric indices in the literature. Here, we 
   will use both $S$ and $\log R'_{\rm HK}$ indices, as defined by Noyes et al. 
   (\cite{noyes}).  
   The first is a measure of the line intensity in the  
   H and K Ca II related to the continuum (Baliunas et al. \cite{baliunas}). 
   From its definition, $S$ depends upon the stellar colour. Noyes et al. provide 
   equations for the transformation of this into $\log R'_{\rm HK}$, which is 
   a colour-independent index.

   The paper is organized as follows: section 2 describes in detail the data, 
   where we pay special attention 
   to the estimate of errors in age and to the representativeness 
   of our sample. In section 3, the AMR is derived and a comparison with  
   previous relations in the literature is presented. A detailed discussion about the 
   magnitude of the metallicity dispersion is given in section 4. The final 
   conclusions follow in section 5.
   
\section{The sample}

\subsection{Selection criteria}

   The criteria we have followed for the construction of the sample were based on  
   the requirement to have a number, as large as possible, of disk stars for which  
   reliable metallicities and ages could be determined. This was done by taking the 
   common stars in the photometric catalogues of Olsen (\cite{olsen83}, \cite{olsen93}, 
   \cite{olsen94}) and the chromospheric activity surveys of Soderblom (\cite{soder}, 
   hereafter S85) and Henry et al. (\cite{HSDB}, hereafter HSDB). This procedure yielded 
   an initial sample composed by 729 late-type dwarfs with $0.307<(b-y)<0.622$ and 
   $-5.40 < \log R'_{\rm HK} < -3.78$. Here, $(b-y)$, as well as $m_1$ and $c_1$ refer 
   to the standard $uvby$ indices (Crawford \cite{crau}).

   According to HSDB, the late-type dwarfs in the solar neighbourhood can be divided 
   according to four chromospheric populations, namely the very active stars 
   ($\log R'_{\rm HK}$ $\ge -4.20$), the active stars ($-4.20> \log R'_{\rm HK} \ge -4.75$), 
   the inactive stars ($-4.75 > \log R'_{\rm HK} \ge -5.10$) and the very inactive 
   stars ($\log R'_{\rm HK} < -5.10)$. We will refer to these groups, hereafter, as 
   VAS, AS, IS and VIS, respectively.
   
\begin{figure}
      \resizebox{\hsize}{!}{\includegraphics{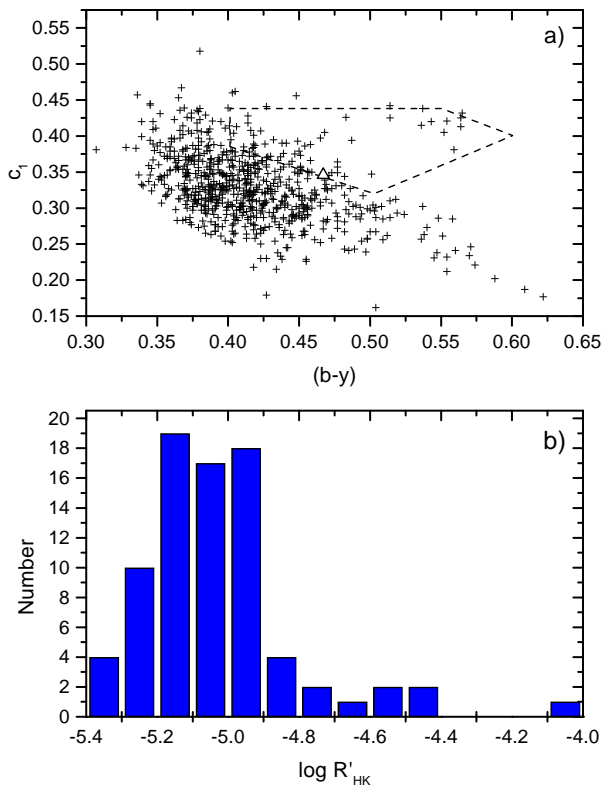}}
      \caption[]{Photometric and chromospheric data for the initial sample of 729 late-type 
      dwarfs. a) Diagram $(b-y)\times c_1$ for the initial sample. The polygon indicates 
      the area populated by subgiants, and the open triangle shows the position of the star 
      HD 119022 in this diagram (see text). b) Distribution of chromospheric activity indices of the 
      stars located in the subgiants area.}
      \label{byc1}
      \end{figure}

   In Figure \ref{byc1}a, the diagram $(b-y)\times c_1$ for these 729 stars is shown. The 
   polygon in this figure corresponds to the area of the diagram occupied mostly by subgiants, 
   according to Olsen (\cite{olsen84}). As much as 11\% of the sample is  
   probably composed of subgiants. Figure \ref{byc1}b shows a histogram of chromospheric 
   activity levels for these stars. Many of them have $\log R'_{\rm HK} < -5.00$, 
   corresponding to a chromospheric age greater than 5.6 Gyr, according to the age 
   calibration by Soderblom et al. (\cite{soder91}), an additional evidence for their 
   evolved status.

   One of these supposed subgiants 
   (\object{HD 119022}, shown in Figure \ref{byc1}a as an open triangle) is a VAS 
   ($\log R'_{\rm HK} = -4.03$), and presumably should be a very young star, not a 
   subgiant. Note that the star is quite near the limit of the subgiant area, and could rather 
   be considered a slightly evolved dwarf. Its nature is still uncertain, 
   since it is located far beyond the zero-age main sequence. Soderblom et al. (\cite{SKH}) 
   have studied this star in more detail, and have found spectral features typical of 
   youth (strong Li and broad-line 
   spectrum), but its high luminosity prevented them from classifying it unambiguously 
   as a young star. Other trends in its spectrum (strong H$\alpha$ in absorption, a discrepancy 
   between colour and spectral type, and deep sharp features inside the 
   Na D profiles) suggest that it could be a heavily reddened star. We will return to this 
   star later, and show that the reddening does affect it substantially.

   These `subgiant' stars are probably a mix of thick disk and old thin disk stars. 
   We choose to keep all 
   of them in the initial sample since their number 
   is relatively small and because they most probably represent 
   the older epochs of the disk evolution we are trying to recover. However, it is not presently 
   known whether their chromospheric properties can be related to that of main-sequence stars. 
   At least, as will be demonstrated in the next pages, there is good agreement between 
   their isochrone ages and chromospheric ages.

 \subsection{Metallicities and colour indices}
   
   The calibrations of Schuster \& Nissen (\cite{schuniss}) were used for the 
   determination of the metallicity of each star. The metallicities for 3 stars 
   redder than $(b-y)=0.599$ were given by the calibration of Olsen (\cite{olsen84}) 
   for K2-M2 stars. To obtain the colours $\delta m_1$ and $\delta c_1$, we 
   have adopted the standard curves $(b-y)\times m_1$ and $(b-y)\times c_1$ given by 
   Crawford (\cite{crau}) for late F and G0 
   dwarfs, and by Olsen (\cite{olsen84}) for mid and late G dwarfs. The error in the 
   metallicity from these calibrations is expected to be around 0.16 dex (Schuster 
   \& Nissen \cite{schuniss}).
  
   To account for the $m_1$ deficiency in active stars (Gim\'e\-nez et al. \cite{gimenez}, 
   and references therein), the photometric metallicity of the active and very active stars 
   ($\log R'_{\rm HK} \ge -4.75$) needs to be corrected by adding to it an amount 
   $\Delta{\rm [Fe/H]}$, 
   given by the equation proposed by Rocha-Pinto \& Maciel (\cite{RPM98}):
   \begin{equation}
   \Delta{\rm [Fe/H]}=2.613+0.550\log R'_{\rm HK}.
   \label{delfeh}
   \end{equation}

   Since this correction is still rather uncertain, we will always present  
   the uncorrected metallicity (which is the quantity that comes directly 
   from the photometric data) in all plots, 
   tables and references in the text, except when mentioned otherwise. 

 \subsection{Chromospheric ages}

   Chromospheric ages were calculated using the age calibration developed by Soderblom 
   et al. (\cite{soder91}; their Equation 3). Besides this calibration, we have tested 
   the calibration by Donahue (1993, as quoted by HSDB). This last calibration 
   gives ages very 
   similar to those of Soderblom 
   et al. (\cite{soder91}), except for stars with $\log R'_{\rm HK} > -4.40$, for which 
   it predicts ages systematically lower. We have not used it in our final results for 
   the sake of consistency, as some corrections applied to the chromospheric ages were 
   based on the calibration by Soderblom et al. (\cite{soder91}). Moreover, only 6\% of 
   the sample stars have $\log R'_{\rm HK} > -4.40$. We have also verified that our 
   conclusions are not dependent on the difference between 
   these age calibrations.

      \begin{figure}
      \resizebox{\hsize}{!}{\includegraphics{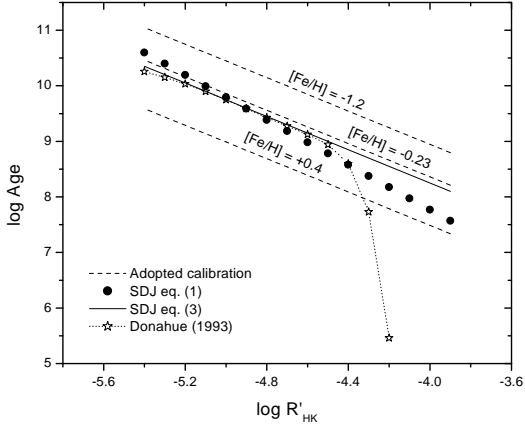}}
      \caption[]{Comparison between chromospheric age calibrations in the literature. Solid 
        line, Soderblom et al. (\cite{soder91}, equation 3); circles, 
        Soderblom et al. (\cite{soder91}, equation 1); stars, Donahue (\cite{don});
        dashed line, adopted calibration for three metallicities, namely 
        $-1.2$, $-0.23$ and $+0.4$.}
      \label{calib}
      \end{figure}

      \begin{figure}
      \resizebox{\hsize}{!}{\includegraphics{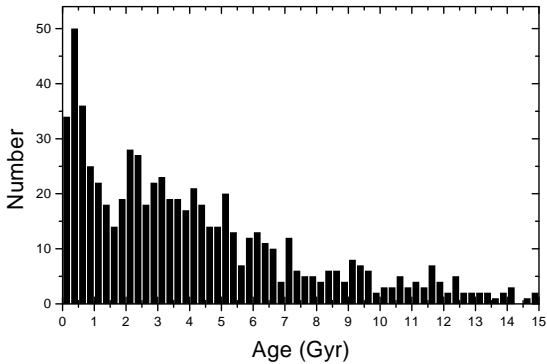}}
      \caption[]{Age distribution for the initial sample. Stars with chromospheric ages 
       greater than 15 Gyr are not shown in this plot.}
      \label{agedist}
      \end{figure}

   The adopted age calibration was further 
   corrected for the dependence on the metallicity proposed 
   by Rocha-Pinto \& Maciel (\cite{RPM98}). In Figure \ref{calib}, we show a comparison 
   between the uncorrected age calibration, and the same curve corrected for three values 
   of [Fe/H], namely $-1.2$, $-0.23$ and $+0.4$, 
   which corresponds to the lowest, the average and the highest metallicity of our sample. Also 
   presented in the same plot are the age calibrations by Soderblom 
   et al. (\cite{soder91}; their Equation 1) and that by Donahue. Most important in this 
   plot is that it shows that the majority of our stars will have an age not very different 
   from that given by 
   the uncorrected Soderblom et al. (\cite{soder91})'s age calibration. This can be 
   seen from the close agreement between the uncorrected age calibration and that for the 
   average metallicity of the sample. 

   Figure \ref{agedist} shows the resulting age 
   distribution for all these stars, excluding 54 stars that present ages 
   greater than 15 Gyr. The unrealistically high age of these 54 stars could be caused by 
   one of these reasons: (i) the star is experiencing a Maunder-minimum phase (see Baliunas et al. 
   \cite{baliunas}); (ii) the errors in the chromospheric index and in [Fe/H] worked 
   together to produce a spuriosly high age; and (iii) the age calibration could be 
   overestimated for large $\log R'_{\rm HK}$ values, since at that range it is given by 
   an extrapolation of data for younger stars. 

      \begin{figure*}
      \resizebox{\hsize}{!}{\includegraphics{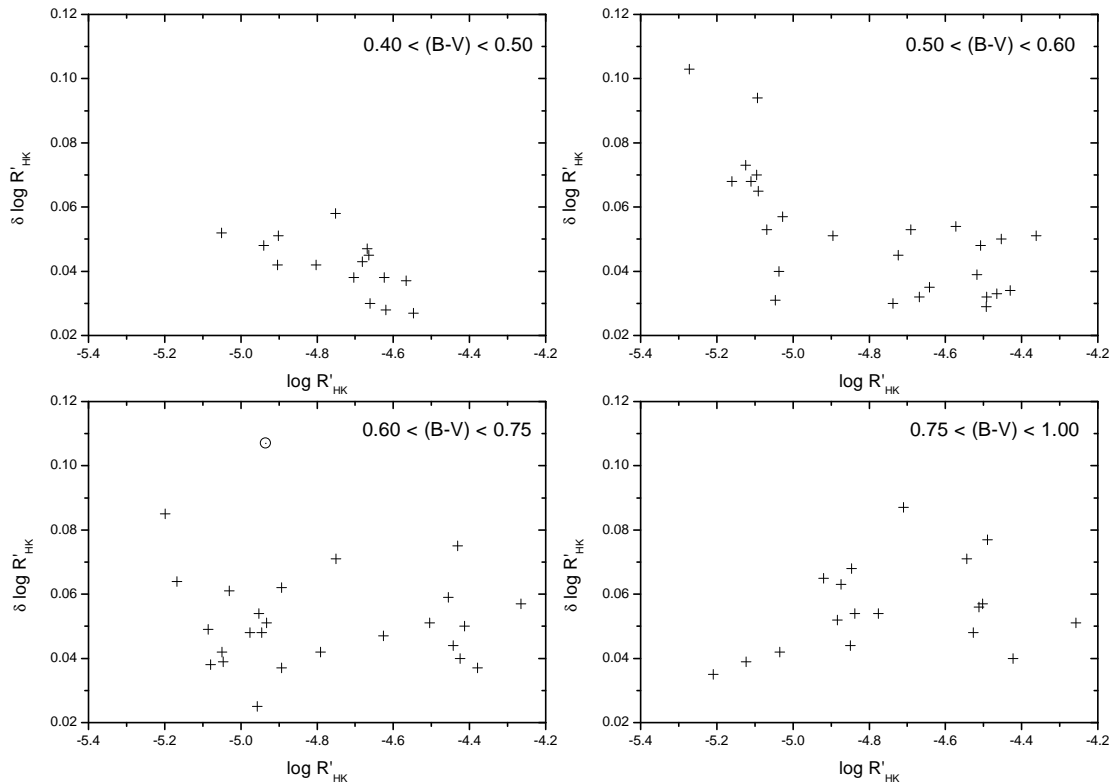}}
      \caption[]{Error in the index $\log R'_{\rm HK}$ due to the stellar magnetic cycles, 
        calculated from the data for 85 stars published by Duncan et al. (\cite{duncan}). A 
        small trend is seen in the sense to find the greatest errors in the old redder dwarfs, 
        but the behaviour is not so well defined to allow the use of a formalism describing it. 
        The Sun is also shown in this diagram, in the bottom-left panel. }
      \label{errcycle}
      \end{figure*}

   Several 
   errors contribute to the error in age, coming from the procedures of corrections and 
   transformation of indices into age. We consider two error sources, namely the 
   error in the index $\log R'_{\rm HK}$ and the error in the photometric metallicity 
   which enters in the metallicity-dependent correction. Rigorously speaking we should also 
   consider the error in Soderblom et al. (\cite{soder91})'s calibration and  
   the error in the metallicity correction itself. However, these are not independent 
   from the former, since the scatter in the calibration reflects mainly the 
   neglecting of the metallicity correction, as well as the uncertainty in the index.

   The average error in $\log R'_{\rm HK}$ was estimated 
   from the data published by Duncan et al. (\cite{duncan}). 
   They present data for the variation of the $S$ index in several late-type stars, 
   for a time interval of 17 years. 
   We have calculated the corresponding values for 
   $\log R'_{\rm HK}$ using the equations by Noyes et al. (\cite{noyes}). The index shows 
   much variation for some stars, but the average error is around $\pm 0.05$. This error estimate 
   agrees closely with that made by S85. 
   The uncertainty introduced in the age by the photometric metallicity 
   was estimated by propagating the error in [Fe/H], using the metallicity-corrected age 
   calibration given by Rocha-Pinto \& Maciel (1998).

   A further complication arises if we consider other error sources: flaring, rotational modulation 
   of active regions, and the stellar 
   magnetic cycles. A small part of this error is already incorporated into the error in 
   $\log R'_{\rm HK}$, but some stars show much variation in the 
   chromospheric indices. Donahue (\cite{don98}) shows that the age of the Sun could be 
   miscalculated within an 
   error of several Gyr if its chromospheric age were derived from observations in an 
   epoch of maximum or minimum activity. S85 remarks that the most serious problem is with the stellar 
   magnetic cycles, and the effects of the other sources of variability are smaller. 
   This uncertainty is the major problem of the 
   chromospheric ages. It can be much decreased provided that we have a good 
   determination of the mean stellar activity, which is still not the case for the majority 
   of the stars in the solar neighbourhood. The promising accuracy of this technique 
   was presented also by Donahue (\cite{don98}), who has shown that the age discrepancy 
   between binaries is usually lower than 0.5 Gyr for systems younger than 2 Gyr, and does 
   not exceed 1.0 Gyr for older pairs.

   We have investigated this error with more detail in the data 
   published by Duncan et al. (\cite{duncan}), for 85 stars that have been 
   continuously monitored from 1966 to 1983. The $\log R'_{\rm HK}$ index for each of 
   these stars was calculated, and an average error was found. The results are presented in 
   Figure \ref{errcycle}, where we have separated the stars according to their colours. 
   The Sun is also shown in this diagram (in the bottom-left panel). To calculate its position, we 
   have used the $S$ indices for the maximum and minimum activity during 
   cycles 20-22, given by Donahue (\cite{don98}).

   Note that there is a small tendency to find the biggest errors in 
   the redder stars. This pattern follows more or less closely 
   the findings by Baliunas et al. (\cite{baliunas}), who have shown  
   that F dwarfs do not show much magnetic variability, while the redder dwarfs present 
   very well defined magnetic cycles. This can increase the error in $\log R'_{\rm HK}$ 
   beyond the value we have adopted, since from only one observation it is impossible to 
   know exactly in what part of the cycle the star is. On the other hand, it is evident 
   from the figure that no single law can be used to estimate the error in the index, given 
   the chromospheric activity level and the stellar colour. 

   The $\log R'_{\rm HK}$ error for the Sun is one of the greatest. It is not clear 
   yet whether it is caused by our more accurate knowledge of the solar magnetic activity 
   (for instance, the time span for the solar $S$ 
   measurements in this figure is about 3 times longer than that of the stars), or to a 
   real more pronounced magnetic variability in the Sun.

   Around 80\% of our sample is composed of stars having only one $\log R'_{\rm HK}$ measure. 
   Their ages are most subject to the errors due to stellar magnetic cycles. Nevertheless, 
   the incorporation of this kind of error is very difficult, and we have decided to use a 
   conservative value of 0.05 dex calculated as described above. This error is 0.01 dex greater than 
   that estimated by S85 for 33 dwarfs. According to this author, on the average the star will present an 
   uncertainty around 10\% in $\log R'_{\rm HK}$ (0.04 dex), due to the stellar magnetic cycles. 
   This procedure will not 
   affect significantly our conclusions, since the error in $\log R'_{\rm HK}$ is small 
   compared to that due to the photometric metallicity.

\begin{figure}
      \resizebox{\hsize}{!}{\includegraphics{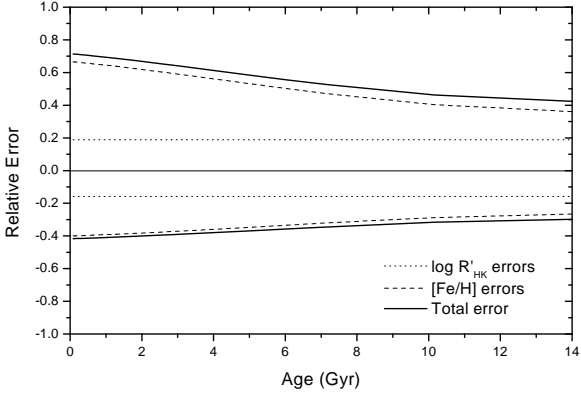}}
      \caption[]{Relative errors in the chromospheric ages. The curves show the individual 
       contribution of each error sources to the error in age. The curves stand for 
       the error in the index $\log R'_{\rm HK}$ (dotted lines), and in the photometric 
       metallicity (dashed lines). The final age error was calculated by the root mean 
       square of all these errors and is shown as 
       thick solid lines.}
      \label{ageerrors}
      \end{figure}

    The impact of the individual error sources on the age is shown in Figure \ref{ageerrors}, 
    which shows relative errors ($\varepsilon_{\rm age}/{\rm age}$), as a function of age. 
    The final error in age, shown as a solid line, was 
    calculated by adding those two individual errors in quadrature. This is the error 
    estimate used throughout this paper.  
    As can be seen from the error trends in the plot, the positive error in the age 
    is greater than the negative error. This is due to the logarithmic nature of the age 
    calibration and the metallicity corrections. We expect that many stars will show ages 
    scattered over a large range of values above the real stellar age. This is one of the 
    reasons why some 
    stars present unreasonable ages greater than 15 Gyr.

\begin{figure}
      \resizebox{\hsize}{!}{\includegraphics{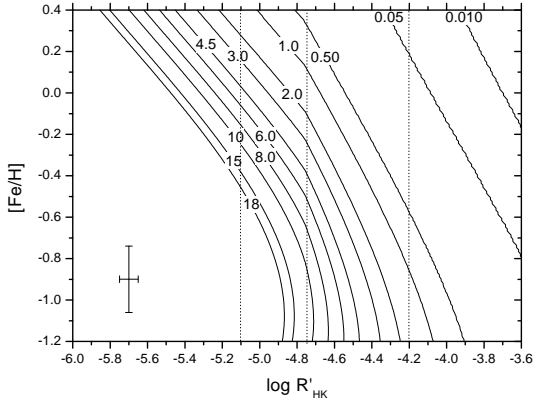}}
      \caption[]{Curves of equal chromospheric age in the metallicity--activity 
         diagram. The average error in this diagram is shown at the lower left corner. 
         The vertical dotted lines separate the four groups of stellar activity, as 
         defined in section 2.}
      \label{chromiso}
      \end{figure}

    Since the chromospheric age depends not only on the activity level of the star but 
    also on its metallicity, the task of finding a chromospheric age is analogous to 
    find an `isochrone' age in the metallicity--activity diagram, as shown in Figure 
    \ref{chromiso}. This figure shows lines of equal chromospheric age, 
    expressed in Gyr. The dotted 
    vertical lines are used to separate the four chromospheric populations, as defined 
    section 2.1. An interesting result is that while for G dwarfs 
    the younger isochrones are crowded in the colour--magnitude diagram, preventing 
    one from obtaining accurate isochrone ages for very young stars, they are much more 
    spaced in the metallicity--activity diagram. The opposite occurs for the 
    older isochrones being more crowded in this last diagram. From this we can see that 
    chromospheric ages are most useful for younger stars, losing part of their accuracy 
    as we consider older stars.

     \begin{figure}
      \resizebox{\hsize}{!}{\includegraphics{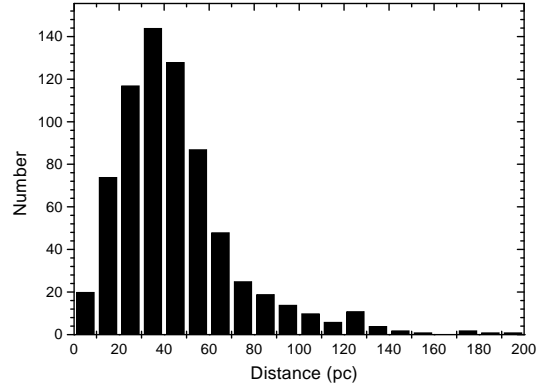}}
      \caption[]{Histogram of distances from the Sun, for the initial sample. The majority 
        of the stars are located within 60 pc, but 5\% of them have distances greater than 100 pc.}
      \label{histdist}
      \end{figure}

  \subsection{Parallaxes, absolute magnitudes and reddening}

    Parallaxes and absolute magnitudes were obtained from the HIPPARCOS data for 714 
    of these stars. The histogram in Figure \ref{histdist} shows the distribution of 
    distances of the stars from the Sun. The majority of the stars are located 
    within 60 pc, but 5\% of them have distances greater 
    than 100 pc.
    
    We have calculated the reddening for 13 of these stars located beyond 100 pc, using 
    $\beta$ indices from Hauck \& Mermilliod (\cite{HM98}) and the intrinsic colour 
    calibration of Schuster \& Nissen (\cite{schuniss}). The results are presented in Table 
    \ref{red} where we list the HD number, the distance from the Sun, 
    $\log R'_{\rm HK}$, the colour excess $E(b-y)$, 
    the dereddened 
    Str\"omgren indices $(b-y)_0$, $m_0$ and $c_0$, the photometric metallicity calculated 
    with the dereddened indices, and the difference between the previous [Fe/H] calculated 
    without and with reddening correction. Four of these stars present $E(b-y)<0$ and should 
    not be reddened. Thus, dereddened indices are not provided for them in the Table.

    The Table 
    shows that there is considerable reddening for some stars, and this can seriously affect 
    the stellar metallicity. Provided that we had $\beta$ for all the stars located beyond 
    100 pc, we could deredden their indices and find more reliable estimates for 
    [Fe/H], and consequently age, but this index is available only for a few stars. To avoid 
    reddened stars, in the next sections we will consider generally only those stars 
    located within 80 pc, 
    for which reddening is expected to be negligible (Olsen \cite{olsen84}).

     Note that the reddening for \object{HD 119022} 
     is appreciable, confirming the suggestion by Soderblom et al. (\cite{SKH}). Moreover, 
     taking the dereddened colours in Table \ref{red}, we see that the star resides  
     outside 
     the subgiant polygon in Figure \ref{byc1}a, right in the bulk of main-sequence stars. 
     This can be taken as additional evidence for a lower age.  

     \begin{table*}
      \caption[]{Data for 13 reddened stars in the sample.}
         \label{red}
         \begin{flushleft}
    {\halign{%
    \hfil#\hfil & \qquad\hfil$#$\hfil & \qquad\hfil$#$\hfil & \qquad\hfil$#$ & \qquad\hfil$#$\hfil  & 
    \qquad\hfil$#$\hfil & \qquad\hfil$#$\hfil  & \qquad\hfil$#$\hfil & 
    \qquad\hfil$#$\hfil  \cr
    \noalign{\hrule\medskip}
    Name & d({\rm pc})& \log R'_{\rm HK} & E(b-y) & (b-y)_0 & m_0 & c_0 & {\rm [Fe/H]}_{\rm 
    dered} & \Delta {\rm [Fe/H]} \cr
    \noalign{\medskip\hrule\medskip}
         \object{HD 3611} & 104.6 & -5.150 & 0.014 & 0.374 & 0.166 & 0.404 & -0.386 & -0.030 \cr
         \object{HD 17169} & 107.9 & -4.720 & 0.040 & 0.432 & 0.231 & 0.332 & -0.044 & -0.300 \cr
         \object{HD 39917} & 194.9 & -4.050 & 0.064 & 0.454 & 0.254 & 0.298 & -0.180 & -0.444 \cr
         \object{HD 119022} & 124.2 & -4.030 & 0.054 & 0.420 & 0.231 & 0.336 & +0.053 & -0.397 \cr
         \object{HD 122683} & 176.7 & -4.760 & -0.018 & & & & & \cr
         \object{HD 139503} & 109.9 & -5.180 & 0.017 & 0.376 & 0.182 & 0.385 & -0.057 & -0.144 \cr
         \object{HD 141885} & 104.1 & -5.340 & 0.019 & 0.390 & 0.208 & 0.459 & +0.155 & -0.160 \cr
         \object{HD 142137} & 105.8 & -5.040 & 0.024 & 0.381 & 0.194 & 0.435 & +0.059 & -0.210 \cr
         \object{HD 151928} & 114.8 & -5.070 & -0.023 & & & & & \cr
         \object{HD 159784} & 124.8 & -5.200 & -0.032 & & & & & \cr
         \object{HD 179814} & 133.3 & -4.970 & -0.017 & & & & & \cr
         \object{HD 199017} & 173.6 & -4.920 & 0.026 & 0.412 & 0.213 & 0.325 & -0.072 & -0.195 \cr
         \object{HD 202707} & 112.0 & -5.110 & 0.027 & 0.396 & 0.185 & 0.343 & -0.188 & -0.225 \cr
  \noalign{\medskip\hrule}}}
         \end{flushleft}
   \end{table*}

     Five objects amongst the 13 VAS of our sample have distances greater than 100 pc. 
     We expect that they are mildly to strongly reddened. The straightforward consequence 
     of this is that their 
     photometric metallicities would seem lower than they really are. This can be seen from 
     Table 1 where there are two VAS listed: \object{HD 39917} and \object{HD 119022}, with 
     a metallicity difference of around $-0.4$ dex. Rocha-Pinto \& Maciel (\cite{RPM98}) 
     showed 
     that the photometric metallicity distribution of the VAS is peculiarly concentrated 
     at lower [Fe/H], in strong contrast with the expected youth of such stars. 
     We have explained this trend as the result of the $m_1$ deficiency, which makes stars 
     with strong chromospheres resemble metal-poor stars from a photometric point of view. 
     This deficiency can originate from the filling in of the line cores due to a 
     chromosphere-driven photospheric activity (Basri, Wilcots \& Stout \cite{basri}). 

     However, these new results suggest that the low photometric metallicities of the VAS 
     could be explained as an effect of the reddening. At least, the difference in the 
     metallicity, $\Delta {\rm [Fe/H]}$, is large enough to make the photometric metallicity of the VAS 
     very similar to the metallicity of the AS, in the ${\rm [Fe/H]}\times \log R'_{\rm 
     HK}$ diagram. We need to know if the reddening can explain the `activity strip' found 
     in this diagram (Rocha-Pinto \& Maciel \cite{RPM98}), since our procedure to correct 
     [Fe/H] in the AS and VAS, for the 
     effects of the $m_1$ deficiency, depends on the meaning of this strip.

     Only one other VAS has a published $\beta$ in the literature. It is \object{HD 123732} 
     which shows a very small reddening, $E(b-y)=0.002$, in agreement with its distance 
     from the Sun (around 63 pc). However, its photometric metallicity is not very low 
     (${\rm [Fe/H]} = -0.220$) compared to other VAS. If we were to trust its youth, we 
     would need to explain its subsolar [Fe/H] as an evidence towards the $m_1$ deficiency. 
     Notwithstanding, Soderblom et al. (\cite{SKH}) have found features in its spectrum that 
     classify it as a W UMa star, so that its chromospheric activity must not come from 
     youth. On the other hand, Soderblom et al. measured spectroscopic [Fe/H] in two 
     VAS which are most probably very young single stars, and the difference between 
     these metallicities and the photometric metallicities are in very good agreement with our 
     prediction in Equation (\ref{delfeh}). 
     Both are located within 50 pc, and therefore are unreddened. The difference between 
     the spectroscopic and photometric [Fe/H] can only be understood as resulting from 
     the $m_1$ deficiency.

\begin{figure}
      \resizebox{\hsize}{!}{\includegraphics{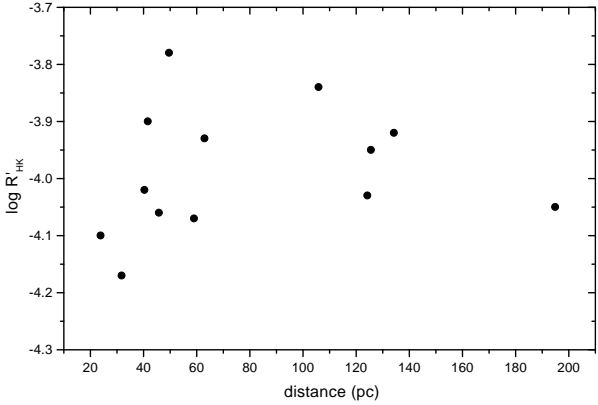}}
      \caption[]{Activity level versus distance from the Sun, for the VAS. The plot shows that a 
     correlation between these quantities probably does not exist. The most distant VAS are not the most active. 
     While reddening can explain part of the low photometric metal-content of some VAS, there is still another 
     effect to account for, and the $m_1$ deficiency seems the most promising one.}
      \label{absurdum}
      \end{figure}

     It is important to bear in mind that what Rocha-Pinto \& Maciel (\cite{RPM98}) have 
     found seems to be a 
     systematic trend depending on $\log R'_{\rm HK}$: the more active the star is, the 
     more metal-deficient it looks. This trend makes the $m_1$ deficiency hypothesis very 
     appealing, since it can produce a similar effect. If this behaviour 
     is likely to be produced instead by reddening, then we need to consider a correlation 
     between distance and $\log R'_{\rm HK}$. Figure \ref{absurdum} shows that such a 
     correlation probably does not exist. The most distant VAS are not the most active. In fact, 
     even disregarding the most distant VAS, the ${\rm [Fe/H]}\times \log R'_{\rm HK}$ 
     diagram still presents the `activity strip'. The average metallicity of the VAS within 
     and beyond 80 pc (our chosen distance cutoff) is $-0.318$ and $-0.480$ dex, respectively, 
     showing that while reddening can explain part of the low photometric metal-content of 
     the VAS, there is still another effect to account for, and the $m_1$ deficiency seems 
     the most promising one.

     Since the proposed correction for the $m_1$ deficiency was empirically determined using 
     only AS stars, and they do not seem to be affected by reddening, we have no 
     reason to disregard them. Indeed, the agreement of our proposed correction with the 
     two VAS studied by Soderblom et al. (\cite{SKH}) is good evidence that the extrapolation of 
     the relation for the VAS is reasonable.

  \subsection{Effects of unresolved binarity}

     According to Duquennoy \&  Mayor (\cite{mayor}), 65\% of the G dwarfs in the solar neighbourhood 
     are presently unresolved 
     binaries. They are expected to present colours contaminated by the 
     unresolved secondary, which would introduce errors in their metallicities 
     and ages. We are here primarily concerned in the age errors, that can be 
     more easily estimated.  Since there is no set of chromospheric measurements for 
     combined and resolved binaries, this discussion is predominantly theoretical.

      \begin{figure}
      \resizebox{\hsize}{!}{\includegraphics{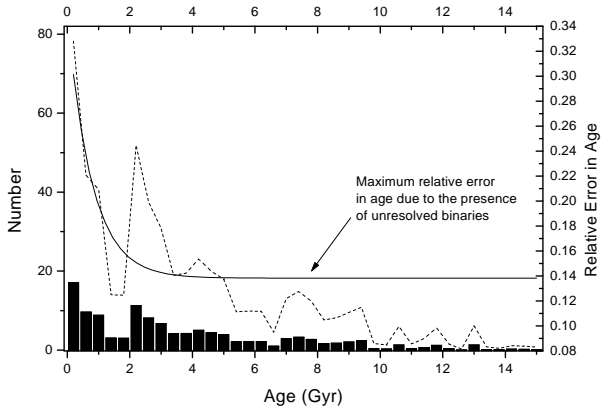}}
      \caption[]{Expected number of binaries in each age bin, shown by dark bars. 
         The dashed line shows the total number of stars in each bin. The solid line 
         indicates the expected relative age error of the unresolved stars, as a function 
         of age. The error is always positive, since unresolved stars appear younger than 
         they really are.}
      \label{binarias}
      \end{figure}

     In principle, for two stars with the same age and metallicity, $\log R'_{\rm HK}$ are 
     equal. The index that depends on 
     the colour is the $S$ index. For two coeval stars, born from the same parental cloud, $S$ 
     will be greater in the cooler one. The index is very dependent on the stellar colour, 
     since the chromospheric flux is much greater in the late type stars. When two stars are 
     observed as one star, the $S$ index of the primary will be contaminated by that of the 
     secondary. The combined index will be always greater than that of the primary, and the 
     star would appear more active and younger than it really is.

     We have calculated the $S$ index that would correspond to a certain $\log R'_{\rm HK}$ 
     for several binary pairs, with $(B-V)$ varying from 0.4 to 1.0, by inverting the 
     equations provided by Noyes et al. (\cite{noyes}). A larger colour range could not be 
     used due to the limit of the $S$ calibration to colours higher than $(B-V)=1.0$. The 
     $(B-V)$ colour 
     was also used to estimate roughly the effective temperature and mass of the star. We thus can 
     calculate the mass ratio for each pair (secondary to primary).

     To combine the $S$ index of the two stars in the pair, we used a weighted mean. The weight is 
     the Planck function integrated from $\lambda\lambda$ 3880 to 4020 \AA, which correspond to the 
     spectral range where the Ca II lines are found. In this way, we can found an average $S$ for the 
     unresolved pair, and the corresponding $\log R'_{\rm HK}$. We compared the difference between 
     this average $\log R'_{\rm HK}$ and the real index, known a priori. Basically, the difference 
     is greater for the active stars, since the $S$ index of the secondary becomes much greater 
     than that of the primary. For the inactive stars, it is lower than 0.04 dex which is smaller 
     than the error for the index presented in Figure \ref{ageerrors}. The difference depends also 
     on the mass ratio, and on the primary mass. A larger mass ratio tends to 
     increase the difference. That means that a pair of stars composed of a G+K dwarf will 
     appear more active than that composed by two G dwarfs. But at a certain mass ratio, around 0.55, the 
     behaviour changes and the difference diminishes. That is, a pair composed of a G+M dwarf 
     would be much less contaminated than that composed of a G+K dwarf. This reflects the fact that 
     the intensity of the spectrum of the hotter star becomes much more important than that of the 
     fainter. The dependence on the primary mass is simpler: the more massive is the primary, 
     the weaker is the contamination by the secondary, and the smaller is the difference in 
     $\log R'_{\rm HK}$. 

     Since these are just rough estimates, we decided to keep the qualitative 
     approach, instead of deriving numerical corrections. We have calculated the number of stars in 
     each age bin that can have a wrong age due to unresolved binaries. This number is estimated as
   \begin{equation}
     N_{\rm b} = \Delta N f_{\rm bin} f_{{m_{< 1.2}}} f_q,
   \label{unbin}
   \end{equation}
     where $\Delta N$ is the number of stars in each age bin; $f_{\rm bin}$ is the fraction of unresolved 
     binaries in the sample, taken as 0.65; $f_{{m_{< 1.2}}}$ is the fraction of stars with masses 
     lower than $1.2 M_\odot$, which are expected to be much influenced to contamination by a secondary, 
     according to our calculations; and $f_q$ is the percentage of mass ratios in which the primary $S$ colour 
     are affected by that of the secondary. From a Salpeter IMF, we estimate that 
     $f_{{m_{< 1.2}}}\approx 0.67$. The fraction $f_q$ is conservatively estimated as 0.50 from our 
     calculations. In this last estimate entered the fact that two stars with mass ratio around 1 will 
     not affect the results: they would have the same index; and stars much fainter will not contaminate 
     the index of the primary.

      \begin{figure*}
      \resizebox{\hsize}{!}{\includegraphics{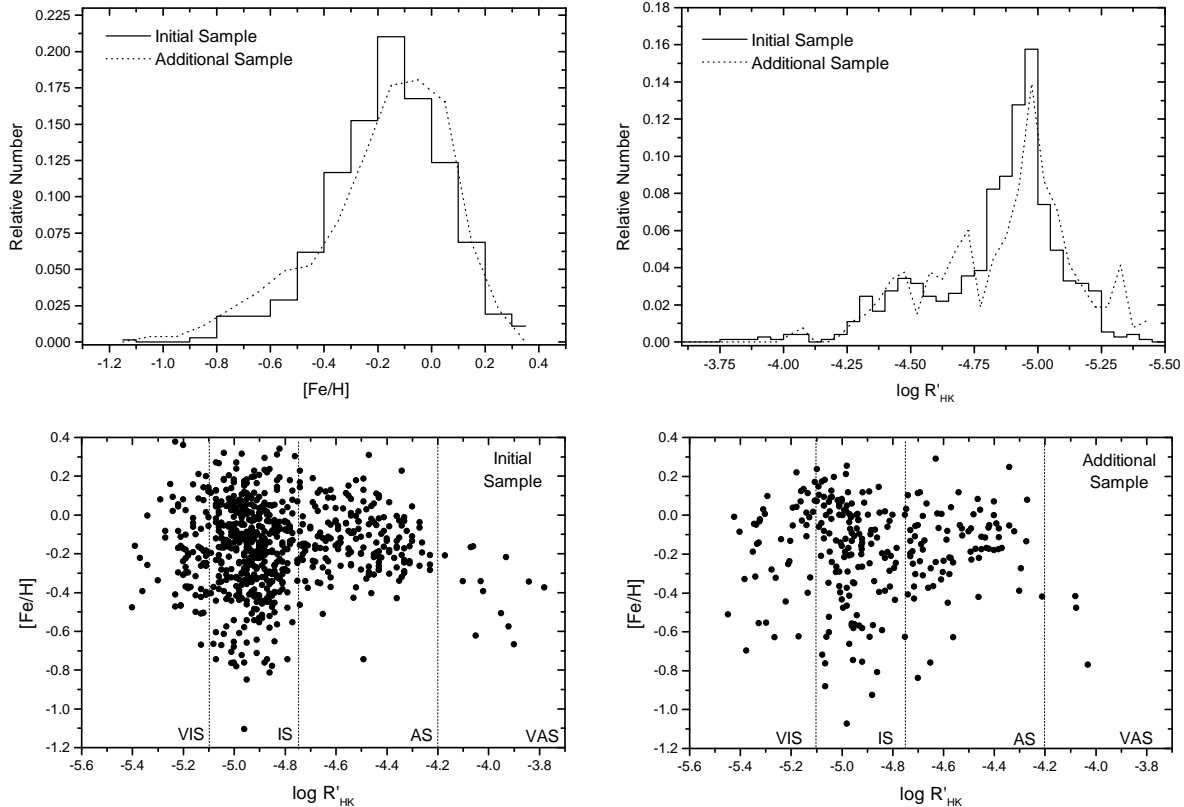}}
      \caption[]{Comparison between the main characteristics of the initial (729 stars) and the 
        additional samples (267 stars). These main characteristics are the metallicity distribution (upper-left 
        panel), chromospheric activity distribution (upper-right panel), and the metallicity--activity distribution 
        (bottom panels).}
      \label{2samples}
      \end{figure*}

     Figure \ref{binarias} shows the number of stars with probably wrong ages due to duplicity. 
     The solid 
     line in this plot gives the expected relative age error for those stars. It uses the 
     maximum error in 
     $\log R'_{\rm HK}$, after varying the primary mass and the mass ratio. The error is always 
     positive, 
     since unresolved stars appear younger than they really are.

     It can be seen that the error is greater for the youngest stars. However, its magnitude 
     is negligible 
     in view of the age errors already considered in Figure \ref{ageerrors}. The number 
     of stars subject 
     to this errors is also small, so that we can conclude that it does not affect the main results 
     of this paper.

  \subsection{Representativeness of the sample}

     An additional sample, consisting of 267 stars, was built in order to complement 
     our initial 
     sample. The sample consists of stars initially in the surveys 
     of S85 and HSDB, but not having photometric data in Olsen's catalogues, as 
     well as other 
     stars scattered amongst several papers by the Mount Wilson group 
     (Soderblom et al. 
     \cite{soder91}; Duncan et al. \cite{duncan}; Baliunas et al. \cite{baliunas}; Saar \& 
     Donahue \cite{saar}), including the Sun itself which has not entered in the initial sample. 
     In some cases, only the $\langle S\rangle$ index was provided, and 
     we calculated the corresponding $\log R'_{\rm HK}$ indices using the formalism of Noyes 
     et al. (\cite{noyes}). Photometric data for these stars were taken from the catalogue of 
     Hauck \& Mermilliod (\cite{HM98}). 

     In figure \ref{2samples} we show the main characteristics of both the initial and the 
     additional sample: the metallicity and $\log R'_{\rm HK}$ distributions and the 
     metallicity--activity diagram. In all three aspects considered, both samples differ 
     somewhat. The metallicity distribution of the additional sample is broader, and the 
     metallicity--activity diagram seems more scattered than that for the initial sample.
     Part of this scatter probably reflects the heterogeneity of the photometric data 
     in the catalogue of Hauck \& Mermilliod (\cite{HM98}). Moreover, there are 
     also significant differences between the chromospheric activity distribution of 
     the two samples. One of these differences is the excess of stars with $-4.75 <
     \log R'_{\rm HK}< -4.60$, where the Vaughan-Preston gap is supposed to be located. Far 
     from being evidence for the non existence of this feature, this excess must be 
     understood as a bias, due to the preferential publication of data for objects with 
     certain activity levels, since they were gathered from scattered papers of the Mount 
     Wilson group aimed at the study of small samples built from different selection 
     criteria. Thus, 
     we cannot join the two samples since the representativeness 
     would be lost. This problem is very important to our study as long as our ultimate 
     goal is counting stars with certain activity levels, after having converted 
     them to ages, to find the star formation history (see Paper II). 
     Since the photometric and chromospheric data come from 
     heterogeneous sources, we decided to use this sample only as an additional tool for 
     our study, in those topics where its inclusion is not likely to change the 
     representativeness of the 
     sample.

      \begin{figure}
      \resizebox{\hsize}{!}{\includegraphics{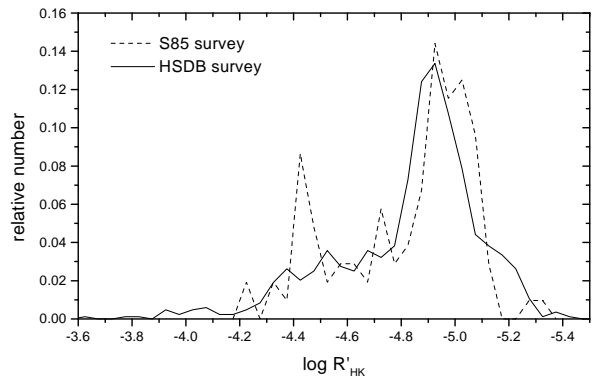}}
      \caption[]{Comparison between the chromospheric activity distributions for the S85 
        and HSDB surveys.}
      \label{2surveys}
      \end{figure}

     In practice, we have to ask whether the initial sample itself is representative 
     since to the present moment the $\log R'_{\rm HK}$ indices for many stars 
     observed by the Mount Wilson group have not been published. We can only base our 
     suppositions about the shape of the stellar chromospheric activity distribution on the 
     two surveys already published: S85 and HSDB. The first of these surveys includes data 
     for 177 nearby stars in the northern hemisphere, while HSDB gives data for 817 southern 
     hemisphere stars. In Figure \ref{2surveys}, we have compared the activity distribution 
     of both surveys. Inspection of this plot shows that the agreement between 
     these distributions is only fair. It is possible to see a 
     separation between active and inactive stars in both surveys, but the relative number 
     of stars in the activity levels seems to be different in them. 

     The HSDB survey has selected stars from the combination of the 
     two-dimensional MK spectral types in the surveys by Houk and collaborators (Houk \& Cowley \cite{HC}; 
     Houk \cite{houk78}; Houk \cite{houk82}; Houk \& Smith-Moore \cite{HSM}) with the photometric 
     data by Olsen (\cite{olsen88}, \cite{olsen93}). A secondary sample composed by 119 stars, not present in 
     Olsen's papers was also observed by them to compensate for the incompleteness of these (see below). 
     
     Houk's surveys includes all stars in the 
     Henry Draper Catalogue, from which Olsen also has constructed his database for photometric observations. 
     The Henry Draper Catalogue is supposed to be nearly complete to magnitude $V<9$, but Olsen's 
     catalogue has some biases towards more massive stars, as explained in Olsen (\cite{olsen93}). 
     However, these biases are not expected to depend upon any age-related quantity. Thus, the 
     only biases expected for our southern hemisphere stars are those present in Olsen's catalogues. The 
     limiting magnitude for completeness is $V=8.3$ mag, which is the brighter cutoff present in 
     Olsen's catalogues (Olsen \cite{olsen83}). 
     For northern hemisphere stars, the sample is based upon the {\it Catalogue of stars within 
     Twenty-five Parsecs of the Sun} (Wooley et al. \cite{wooley}), which also 
     constitutes a complete sample of nearby solar-like stars.

     We are inclined to 
     assume that the HSDB survey best represents the real galactic chromospheric activity 
     distribution, since it includes the largest sample. A comparison between the primary and 
     secondary sample in the HSDB survey (see their Figure 5), for example, shows that the secondary sample 
     (119 stars) present a broader chromospheric acticity distribution between $-4.80 < \log 
     R'_{\rm HK} < -5.10$, agreeing more with the the S85 survey. However, a definitive answer 
     to this question can only be given when data for a sample of northern hemisphere stars, 
     as extensive as that of HSDB, is published. 

\begin{figure*}
      \resizebox{\hsize}{!}{\includegraphics{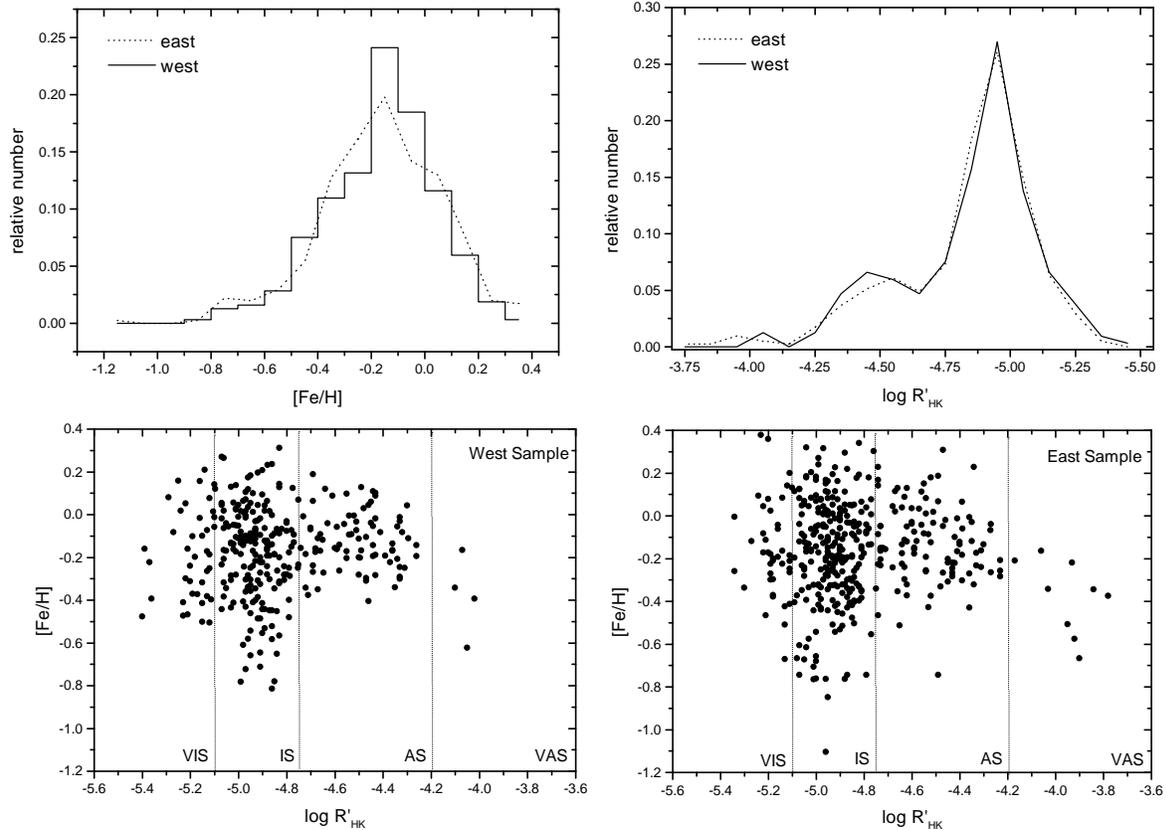}}
      \caption[]{Comparison between the main characteristics of the west (R.A $<12\fh$) and 
       east (R.A $>12\fh$) subsamples. The panels show the same diagrams as in Figure \ref{2samples}.}
      \label{eastwest}
      \end{figure*}

     Around 87\% of our sample is composed of southern hemisphere stars, and this makes 
     our chromospheric activity distribution very similar to that of the HSDB survey. But this 
     does not warrant the representativeness of ours, since we would be comparing 
     very similar samples. At least, we can look for some biases if we divide our sample 
     and verify whether both subsamples keep the same 
     characteristics. We have done this by separating the stars according to right 
     ascension: stars with R.A. from 0$\fh$ to 12$\fh$ compose the west sample (410 stars), 
     while that with R.A. greater than 12$\fh$ compose the east sample (319 stars). In 
     Figure \ref{eastwest} we show the metallicity and $\log R'_{\rm HK}$ distributions and 
     the metallicity--activity diagram for the west and east subsamples. Note that the 
     agreement between them is much better in all three characteristics we have 
     considered. According to this, we can assume that probably there 
     is a representative disk chromospheric activity distribution.

\subsection{Spatial Velocities and Orbital Parameters}
     
     For 460 stars from the initial and the additional sample, a radial velocity measurement
     was found in the literature. These stars compose the `kinematical sample'. 
    Each star had its spatial velocities ($U$, $V$ and $W$) calculated from the data in the literature, 
    using the equations provided by Johnson \& Soderblom (\cite{JS}). 
    Their orbits were then determined by 
    numerical integration within a model of the Galactic 
    potential, for the calculations of the peri- and 
    apo-Galactic distances, $R_p$ and $R_a$ and the mean Galactocentric radius, 
    $R_m = (R_p + R_a)/2$ for the orbit (cf. Edv93), eccentricity and 
    height above the galactic plane. 

     The kinematical sample is particularly used in a future paper of this series aimed to the 
     study of kinematical constraints 
     related to age. Details on this subsample is to be given in that papers.

\section{The chromospheric age--metallicity relation}

\begin{figure}
      \resizebox{\hsize}{!}{\includegraphics{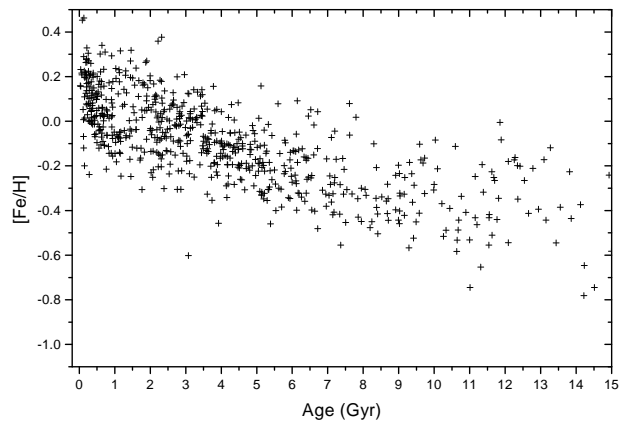}}
      \caption[]{Age--metallicity diagram of the initial sample (729 stars). The metallicities of the VAS 
         and the AS are corrected for the $m_1$ deficiency. Stars with chromospheric ages 
       greater than 15 Gyr are not shown in this plot.}
      \label{rawamr}
      \end{figure}

     For the derivation of the age--metallicity relation, we have to 
     correct the metallicity of the AS and VAS for the $m_1$ deficiency. Figure 
     \ref{rawamr} shows the age--metallicity diagram for the whole initial sample, 
     up to 15 Gyr. 
     The stars seem to fall along a very smooth relation just like what is 
     expected from chemical evolution theory: the rich stars are the youngest, 
     and the poor ones are the oldest. A surprising result is that the scatter is much 
     smaller than that found by Edv93 using isochrone 
     ages. The significance of this will be discussed later.

     To find a more refined and unbiased AMR, the sample needs further selection. 
     First, we have eliminated 72 stars more distant than 80 pc. Taking into 
     account also 11 stars that do not have parallaxes measured by HIPPARCOS, our 
     sample is reduced to 645 stars. As we want to apply a volume correction to 
     the AMR (Twarog \cite{twar}), the sample was further limited to apparent $V$ 
     magnitudes lower than 8.3 mag, by the elimination of 42 stars.

     Eight remaining VAS were also eliminated, since many of them may be old 
     close binaries, in which the high 
     activity is produced by highly synchronized rotation, and is not related with 
     age. This can explain why some stars with ages lower than 0.5 Gyr still present 
     subsolar [Fe/H] in Figure \ref{rawamr}. Rigorously speaking, one may ask why we 
     have not discarded the VIS since 
     their very low activity might also be 
     unconnected with age and could reflect an evolutionary phase analogous to the 
     Maunder Minimum through which the Sun passed during the 17th-18th centuries 
     (Baliunas et al. \cite{baliunas}; HSDB). We keep the VIS in our sample as 
     the effects of the metallicity on the $\log R'_{\rm HK}$ index make the 
     richer stars resemble older IS,  
     or VIS. Many stars amongst the VIS can be 
     normal stars, and there is presently no way to separate them from the Maunder-mininum 
     stars. The same kind of contamination by normal stars does not significantly affect 
     the VAS group, as can be seen from Figure 6 of Rocha-Pinto \& Maciel (\cite{RPM98}). 

     Seven active stars are known to present greater velocity components (Soderblom 
     \cite{soder90}), and they probably are not young stars (Rocha-Pinto, Castilho \& 
     Maciel \cite{RPCM}). All of these stars were eliminated from the sample. 
     Finally, disregarding 37 stars older than 15 Gyr, we arrive at a sample with 
     552 stars. The criteria for elimination are summarized in Table \ref{cut}.

     \begin{table*}
      \caption[]{Criteria for eliminating stars of the sample.}
         \label{cut}
         \begin{flushleft}
    {\halign{%
    #\hfil & \qquad\hfil#\hfil \cr
    \noalign{\hrule\medskip}
    Criterion & stars eliminated \cr
    \noalign{\medskip\hrule\medskip}
         Stars not having parallaxes in the HIPPARCOS database & 11 \cr
         Stars distant by more than 80 pc & 72 \cr
         Apparent V magnitude greater than 8.3 mag & 42 \cr
         Very active stars ($\log R'_{\rm HK} \ge -4.20$) & 8 \cr
         Objects known to be chromospherically young, but kinematically old & 7 \cr
         Stars with nominal chromospheric ages greater than 15 Gyr & 37 \cr
    \noalign{\medskip\hrule}}}
         \end{flushleft}
   \end{table*}

     \subsection{Magnitude-limited AMR}

     Since our sample is not volume-limited, we have to apply a correction to account 
     for the dependence on [Fe/H] of the apparent magnitudes. This procedure, called 
     volume correction, was already used by some authors (Twarog \cite{twar}; Ann 
     \& Kang \cite{annkang}; Meusinger 
     et al. \cite{meu}). After binning the stars, each metallicity is 
     weighted by $d^{-3}$, where $d$ is the maximum distance at which the star would 
     still have an apparent magnitude lower than the magnitude limit, which for our 
     sample is 8.3 mag. It is given by
    \begin{equation}
    8.3-M=5\log d - 5,
   \label{d}
   \end{equation}
     where $M$ is the absolute magnitude of the star.

     \begin{table}
      \caption[]{Magnitude-limited AMR.}
         \label{amr8.3}
         \begin{flushleft}
    {\halign{%
    \hfil#\hfil & \qquad\hfil$#$\hfil & \qquad\hfil$#$\hfil & \qquad\hfil$#$\hfil  \cr
    \noalign{\hrule\medskip}
    Age & N & \langle{\rm [Fe/H]}\rangle & \sigma_{\rm [Fe/H]} \cr
    \noalign{\medskip\hrule\medskip}
         0-1 & 107 & +0.12 & 0.13 \cr
         1-2 & 61 & +0.01 & 0.14 \cr
         2-3 & 79 & -0.08 & 0.12 \cr
         3-4 & 66 & -0.10 & 0.12 \cr
         4-5 & 58 & -0.15 & 0.08 \cr
         5-6 & 44 & -0.26 & 0.11 \cr
         6-7 & 32 & -0.28 & 0.13 \cr
         7-8 & 26 & -0.33 & 0.14 \cr
         8-9 & 17 & -0.36 & 0.10 \cr
         9-10 & 21 & -0.39 & 0.12 \cr
         10-11 & 10 & -0.40 & 0.14 \cr
         11-12 & 13 & (-0.45) & 0.17 \cr
         12-15 & 18 & (-0.48) & 0.20 \cr
  \noalign{\medskip\hrule}}}
         \end{flushleft}
   \end{table}

      \begin{figure}
      \resizebox{\hsize}{!}{\includegraphics{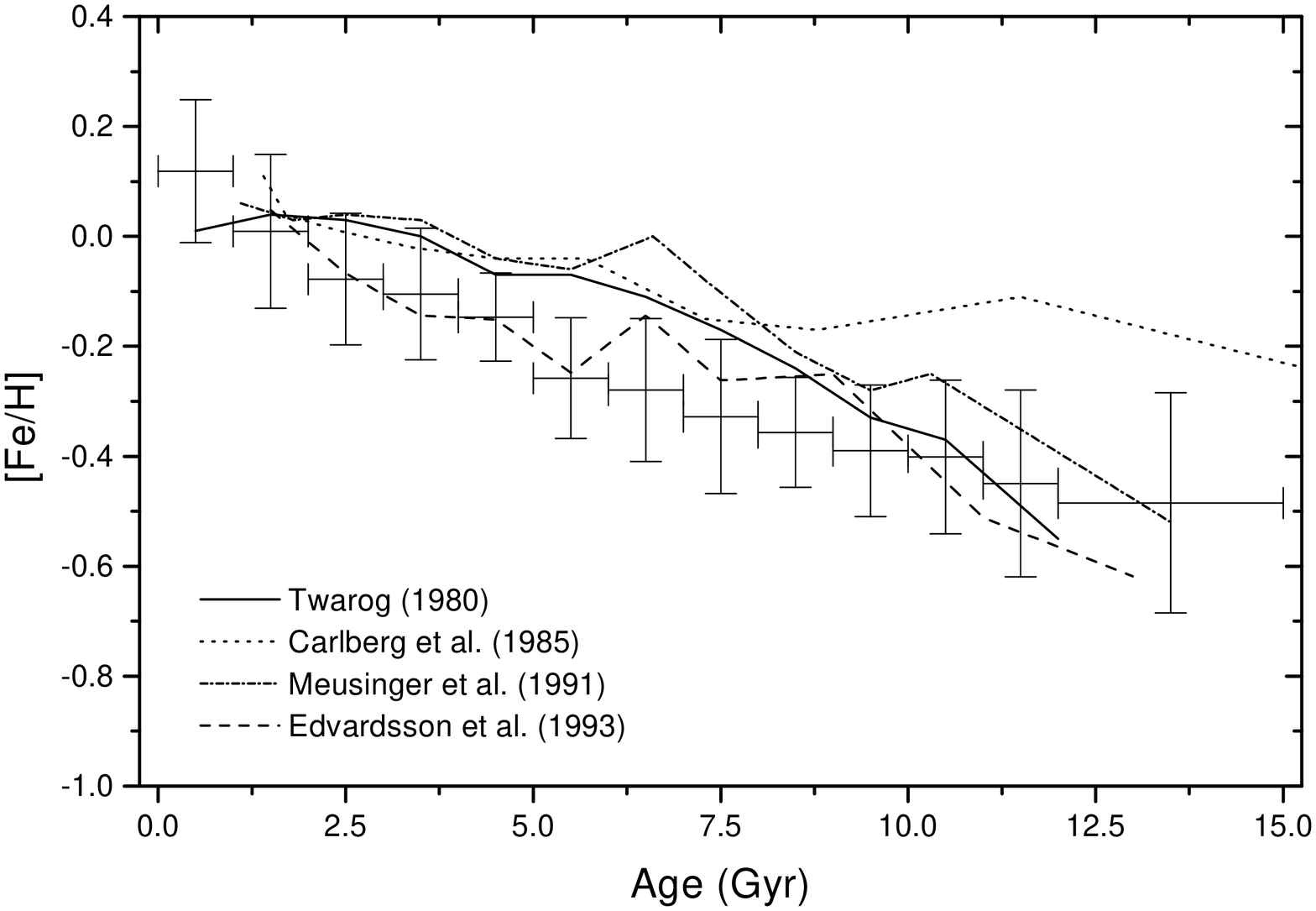}}
      \caption[]{Averaged magnitude-limited AMR (552 stars) compared to previous relations in the literature. 
         The metallicities of the VAS 
         and the AS are corrected for the $m_1$ deficiency}
      \label{amr}
      \end{figure}

     This magnitude-limited AMR is presented in 
     Table \ref{amr8.3}, where we give the number of 
     stars in each metallicity bin, the average-weighted [Fe/H], corrected 
     for the $m_1$ deficiency (see section 2.2), and the metallicity 
     dispersion. The metallicity of the last two bins is 
     presented between parenthesis, since they are 
     expected to be upper limits (see next section). 
     For the calculation of the metallicity dispersion, we have used the same 
     weights, given by Equation (\ref{d}). These data are presented also in Figure \ref{amr}, 
     compared 
     with previous relations published in the literature. In spite of having found a somewhat 
     lower metallicity dispersion, we see a good agreement with the mean points of the 
     Edv93's AMR. The agreement with the AMRs by Twarog (\cite{twar}) and 
     Meusinger 
     et al. (\cite{meu}) is marginal. These AMRs predict a steady growth of [Fe/H] with 
     time, but with a significant flattening in the last 5 Gyr. Instead, we have found a 
     steepening in the growth of [Fe/H] at that same epoch, a feature also apparent in 
     Edv93. In the oldest age bin, our AMR gives an average 
     metallicity 
     of $-0.48$ dex which is $0.14$ dex higher than the average metallicity in 
     Edv93. We believe that this discrepancy results from the 
     errors in the chromospheric ages (see discussion on section 3.2). 

     We have not found an absence of metal-rich stars with ages 
     between 3 and 5 Gyr (Carraro et al. \cite{carraro}). 
     In fact, our metal-rich stars are very concentrated at the 
     younger bins, since this AMR does not show the large scatter present in other studies.

     \subsection{The initial metallicity of the disk}

     The AMR presented in Table \ref{amr8.3} indicates a high estimate for the initial 
     metallicity of the disk. According to it, we should expect an average [Fe/H] 
     $\approx -0.48$ dex at around 13.5 Gyr ago. This metallicity is somewhat higher than 
     the corresponding values found by Twarog (\cite{twar}) and Edv93.

     A high initial metallicity is indicative of significant pre-enrichment 
     of the gas before the formation of the first stars in the disk. That the disk 
     has had some previous enrichment can be easily seen from the lower 
     cuttoff in its metallicity distribution in [Fe/H] $\approx -0.8$ dex (Rocha-Pinto 
     \& Maciel \cite{RPM96}). Even in the framework of an infall model, the disk initial 
     metallicity must be non-zero in order to match the G-dwarf problem.

     Our AMR has a peculiar behaviour towards greater ages. It flattens while 
     the other relations (as well as the theoretical models) generally becomes steeper,
     indicating a rapid enrichment in the early galactic phases. That flatenning 
     is the major difference between the mean points of our AMR and the mean points 
     of that by Edv93.

     The behaviour of our AMR in the oldest age bins are probably very affected by 
     the age errors discussed in section 2. The chromospheric activity 
     yields more accurate ages for younger stars, and we expect that the AMR at 
     the youngest bins is fairly well reproduced. For the older bins, however, we have 
     to estimate the mean deviation from the real AMR that we expect to find from 
     using chromospheric ages. This can be done by simulating the scattering of the 
     stars in the AMR due to the age errors. 

     The procedure follows closely that used in Paper II to compute the statistical 
     confidence levels for the star formation history (hereafter SFH) features, and we 
     refer to that paper for more detail. We simulate a constant SFH composed 
     by 3000 stars. For each star, a metallicity is assigned by a pre-adopted AMR. 
     After that, the stellar age and its metallicity are 
     used to derive the corresponding $\log R'_{\rm HK}$ index the star would 
     present if it is not in a Maunder-minimum phase. This is done by inverting the 
     equations presented by Rocha-Pinto \& Maciel (\cite{RPM98}). A database 
     composed by 3000 stars with $\log R'_{\rm HK}$ and [Fe/H] is then built. The stars 
     in this database are binned in 0.2 Gyr intervals, and we discard randomly in each 
     interval the number of stars expected to have died or to have left the galactic 
     plane at that corresponding age (see Paper II). The remaining 
     stars compose a sample  
     of around 600-700 stars. For these stars, we randomly shift their corresponding 
     $\log R'_{\rm HK}$ and [Fe/H] according to the corresponding errors in these 
     quantities. The final catalogue then resembles very much the real data 
     sample used in this study. 

     The final simulated catalogue is used to derive the AMR in the same form we did 
     for the real data. For the sake of simplicity, we assume a simple linear law 
     for the AMR:
   \begin{equation}
   {\rm [Fe/H]}=0.44-0.10 t.
   \label{AMRfake}
   \end{equation}
     where $t$ is the stellar age in Gyr.

      \begin{figure}
      \resizebox{\hsize}{!}{\includegraphics{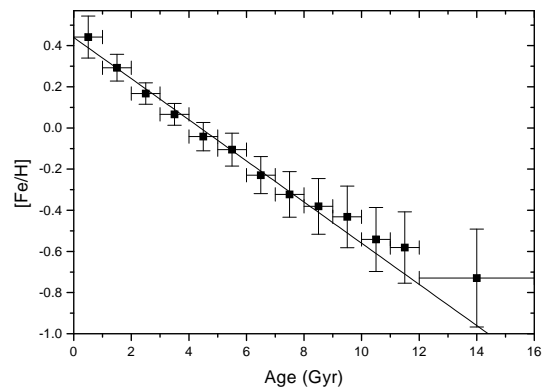}}
      \caption[]{Simulation of the AMR, with a set of 700 stars. The solid curve shows 
          the original linear AMR. The points with error bars indicate the recovered AMR. 
          It can be seen that the chromospheric age errors introduce 
          a small scatter in the AMR, but do not destroy the average relation. }
      \label{amrsimul}
      \end{figure}

     The simulations were repeated 20 times. The results are very similar from one 
     simulation to the other, and one sample simulation is shown in Figure \ref{amrsimul}. It 
     can be seen that the AMR found with the chromospheric ages follows closely 
     the real AMR of the disk up to 9 Gyr ago, when it begins to deviate. The 
     deviation is always in the sense of an increased dispersion and higher  
     mean AMR. This is exactly the same that is observed in the AMR we have found 
     for the disk. On the other hand, these simulations show that even in face of 
     great age errors, the AMR can be fairly well recovered from a sample with 
     chromospheric ages.

     The increase of metallicity dispersion at the older age bins, as well as the 
     greater deviation from the real AMR, are all consequences of the age errors. 
     The older bins tend to be populated by the poorer stars, and the corrections to 
     the chromospheric ages (Rocha-Pinto \& Maciel \cite{RPM98}) increase 
     very much for metal-poor stars. A small error in [Fe/H] or $\log R'_{\rm HK}$ reflects 
     in a substantial error in age, which most probably will push the star to 
     greater chromospheric ages. Due to this effect, the oldest age bins are more likely to 
     be depopulated by metal-poor stars than the other bins. That is why the recovered 
     AMR fails to match the original AMR as we consider ages progressively older. And the 
     metallicity dispersion increase because the absolute age errors increases as 
     a function of age.

     In our simulations we have found that the average metallicity of the oldest bin 
     is invariably around 0.20-0.23 dex higher than the real average metallicity, due 
     to the age errors. If we apply this value to the Milky Way AMR derived previously, 
     we get a disk initial metallicity around $-0.70$ dex, which agrees very well 
     with Edv93's AMR and with the G dwarf metallicity distribution.

  \section{Metallicity dispersion}

     \subsection{Some Evidences for a Non-Homogeneous Interstellar Medium}

     The metallicity dispersion we have found is about 0.13 dex as shown by the 
     error bars of Figure \ref{amr} and from Table \ref{amr8.3}. It is 
     much lower than that found by Edv93. At first sight, this result 
     seems to revitalize old ideas about 
     the chemical evolution of our Galaxy, in which the interstellar medium has been 
     continuously and homogeneously enriched by metals ejected by the stars. Previous 
     works on the AMR (Twarog \cite{twar}; Meusinger et al. \cite{meu}) have consolidated 
     this view by finding $\sigma_{\rm [Fe/H]}$ around 0.12-0.18 dex. 

     This picture was 
     strongly questioned after Edv93 published their detailed 
     work on the chemical evolution of the disk. They found a larger 
     dispersion in the AMR, 
     varying from 0.18 to 0.26 dex. A number of hypotheses to explain it were 
     considered, and the main conclusion of the authors is that a significant part of this 
     scatter could be physical. Infall is quoted as one of the best mechanisms that could 
     drive such an 
     intrinsically large metallicity dispersion at all epochs, if the time scale for 
     the mixing of the infalling gas is greater than that of star formation. The 
     viability of the infall hypothesis requires star formation to be much more 
     efficient in 
     the material that has just fallen onto the disk than in the already well-mixed gas. This 
     would occur if infall could drive star formation. It is interesting to see that 
     a number of theoretical and observational evidences favour this 
     process (L\'epine \& Duvert \cite{lep94}; L\'epine et al. \cite{lep99}), although 
     a detailed study of this star formation mechanism connected with a chemical 
     evolution model has never been made.
     
      \begin{figure}
      \resizebox{\hsize}{!}{\includegraphics{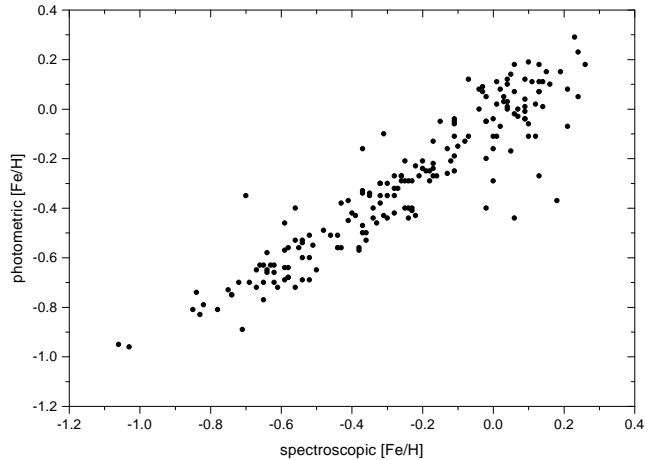}}
      \caption[]{Comparison between photometric and spectroscopic metallicities for Edv93's 
         stars. The photometric metallicity for them was estimated as described in 
         subsection 2.2. }
      \label{metcomp}
      \end{figure}

     An alternative explanation was proposed by Wielen et al. (\cite{WFD}) and Wielen 
     \& Wilson (\cite{wiewil}) according to which the metallicity dispersion in 
     Edv93's AMR originated from the 
     diffusion of the stellar orbits (Wielen \cite{wielen}). According to this hypothesis,
     the Sun was born 1.9 kpc closer to the galactic center in comparison with its present 
     position. This could solve a long-lasting puzzle about the fact that the Sun is richer 
     than some younger neighbour objects (Cunha \& Lambert \cite{cunha}; 
     de Freitas Pacheco \cite{camisa12}). Clayton (\cite{clayton}) also used this scenario 
     to explain the meteoritic abundance ratios of the isotopes $^{29}$Si and $^{30}$Si, related to 
     $^{28}$Si.

     Recently, Binney \& Sellwood (\cite{binney}) reconsidered the diffusion of stellar orbits and 
     found that, a typical star is unlikely to migrate from the galactocentric radius of its 
     birthplace by more than 5\% over its lifetime. These authors say that Wielen et al.  (\cite{WFD}) 
     calculated the diffusion of stellar orbits in velocity space, adopting isotropic and constant 
     diffusion coefficients. When the diffusion is calculated in integral space, it becomes very 
     anisotropic. As a result, the star does not change significantly the guiding-center radius of its orbit, 
     which should be similar to $R_m$, as in Edv93. This recent work concludes that the metallicity scatter 
     in Edv93 is probably real.

     The existence of some scatter in the interstellar medium is not questioned. The Sun-Orion abundance discrepancy  
     and the existence of young stars with subsolar abundance (Grigsby, Mulliss \& Baer \cite{grigsby}) are classical 
     puzzles that point to the existence of an intrinsic metallicity dispersion in the galactic gas. The real  
     problem is its quantification, since these anomalies can be outliers.      

     For instance, Binney \& Sellwood (\cite{binney}) show data measured by other authors, which 
     indicate that the intrinsic 
     scatter in [O/H] should be around 0.10 dex. Garnett \& Kobulnicky (\cite{garnett}) also conclude that, 
     from measurements in nearby galaxies and the local ISM 
     (Kennicutt \& Garnett \cite{KG96}; Kobulnicky \& Skillman \cite{KS96}; Meyer, Jura \& 
     Cardelli \cite{meyer}), the metallicity scatter is lower than 0.15 dex. 

     \subsection{Isochrone-spectroscopic data vs. chromospheric-photometric data}

     The average metallicity dispersion we have found is very close to that found by 
     Twarog (\cite{twar}; $\sim 0.12$ dex). Twarog's sample is greater than Edv93's 
     and has a good statistical significance, but his metallicities are less accurate. 
     Moreover, his ages were found by old, now outdated, isochrones. The problem 
     can be summarized as follows: photometric AMRs show a small, well-behaved 
     metallicity scatter (however, see Marsakov et al. \cite{marsakov}), 
     while the only one based on a spectroscopic sample indicate the opposite. 
     Note that Ng \& Bertelli (\cite{ng}) only revisit Edv93's ages, so that their AMR 
     cannot be taken as an independent evidence for a real greater scatter.

\begin{figure*}
      \resizebox{\hsize}{!}{\includegraphics{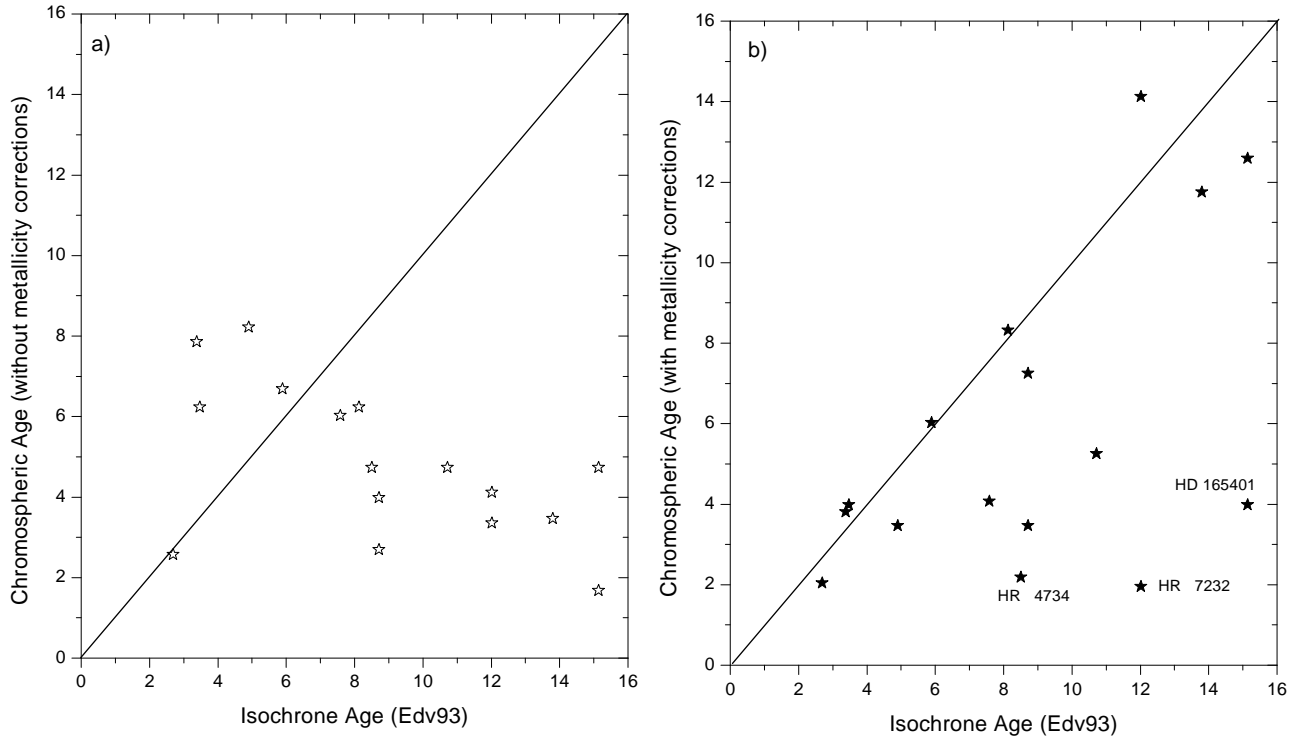}}
      \caption[]{Comparison between isochrone and chromospheric ages for 16 stars in common between this 
         work and Edv93. a) before the application of the metallicity corrections to the chromospheric ages 
         (Rocha-Pinto \& Maciel \cite{RPM98}); b) after the application of the metallicity corrections. The few deviating 
         stars are discussed in section 4.3.} 
      \label{wehoo}
      \end{figure*}

     It is important to show that there is no significant difference in the quality of our data 
     compared to those of Edv93, even taking into account that our metallicities are photometric and 
     our ages are chromospheric. In Figure \ref{wehoo}, we show the comparison between the 16 stars in common 
     between ours and Edv93's sample, before and after the application of the metallicity corrections to the 
     chromospheric ages (Rocha-Pinto \& Maciel \cite{RPM98}). Note that these corrections improve substantially 
     the matching between both methods to measure the stellar ages. The exceptions are few, and will be discussed 
     separately.
     
\begin{figure}
      \resizebox{\hsize}{!}{\includegraphics{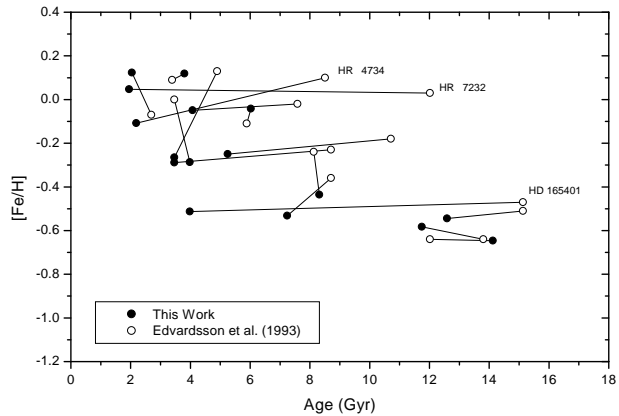}}
      \caption[]{Age--metallicity diagram showing the position of the 16 stars in common between this work and 
         Edv93. The lines connect the same stars as they appear using photometric plus chromospheric data and 
         isochrone plus spectroscopic data. With only one exception, all stars present differences in metallicities 
         that are within the expected error in the photometric calibration of [Fe/H].} 
      \label{amrloci}
      \end{figure}

     The agreement is also good for metallicities. In Figure \ref{amrloci}, the position of our stars in 
     the age--metallicity diagram is shown. Lines connect these stars with their position in the same diagram using 
     Edv93's data. With only one exception, all stars present differences in metallicities that are within the expected 
     error in the photometric calibration of [Fe/H]. 

     Sixteen stars is a small number to test if our method to estimate ages and [Fe/H] is good. We have chosen 
     to compare directly these methods to estimate stellar ages and metallicities, using the stars in Edv93's sample.

     The first of these comparisons is shown in Figure \ref{metcomp}, where we show the 
     correlation of photometric and spectroscopic metallicities for the 189 stars in Edv93 
     sample. The photometric metallicity was estimated as described in the subsection 
     2.2. The agreement is typical of that expected from a photometric calibration. The standard 
     deviation of the data is 0.10 dex.
     
     Chromospheric indices were published only for 81 stars from those of Evd93 sample, and 40 of them 
     were used in the determination of the metallicity correction to the chromospheric age 
     (Rocha-Pinto \& Maciel \cite{RPM98}). We present, in Figure \ref{agecomp}, a comparison 
     between the isochrone and the chromospheric ages, before and after the application of the 
     metallicity corrections. The stars are distinguished by symbol (according to their spectroscopic 
     metallicity) and style (if it was already used by Rocha-Pinto \& Maciel \cite{RPM98} 
     or not). The Figure shows clearly that the metallicity corrections improve substantially the 
     chromospheric ages. The scatter is comparable to the expected error in both methods (we use here 
     a formal average error of 0.1 dex for Edv93 ages, although Lachaume et al. \cite{lach} have shown 
     that this error can be easily underestimated). Panel c of Figure \ref{agecomp} shows the same 
     comparison, now using Ng \& Bertelli (\cite{ng}) 
     ages. The agreement is somewhat better, although a greater isochrone age is 
     still found for some stars. Panel d shows a comparison, in the same scale, between the isochrone age 
     determinations by Edv93 and Ng \& Bertelli (\cite{ng}). 

     Isochrone ages are better compared to chromospheric ages, at least for early G dwarfs (Lachaume et al. 
     \cite{lach} showed that this is not true for intermediate and late G darfs). The difference can be large, 
     but in general, it is of the same order as the difference between two isochrone age determination made using 
     different isochrones (see panel d of Figure \ref{agecomp}). This Figure shows that the error estimate in 
     Fig. \ref{ageerrors} is fairly reasonable. They are considerably large, but they are not expected to destroy the 
     AMR, as shown by our simulation in Figure \ref{amrsimul}.

     There are only three stars that were classified as subgiants in Figure \ref{byc1}, that have 
     both a chromospheric and isochrone age. For two of them (\object{HD 9562} and \object{HD 131117}), 
     the ages agree remarkably to within 0.15 dex. The third star (\object{HR 4734} = HD 108309) is 
     investigated in more detail in what follows. On the other hand, Edv93's sample has some slighlty 
     evolved subgiants, which are represented by the oldest stars in Figure \ref{agecomp}. The good agreement 
     between their chromospheric and isochrone ages shows that the chromospheric activity--age relation 
     can be extrapolated to slightly evolved subgiants.

     \subsection{Anomalies to the chromospheric activity--age relation}

\begin{figure*}
      \resizebox{\hsize}{!}{\includegraphics{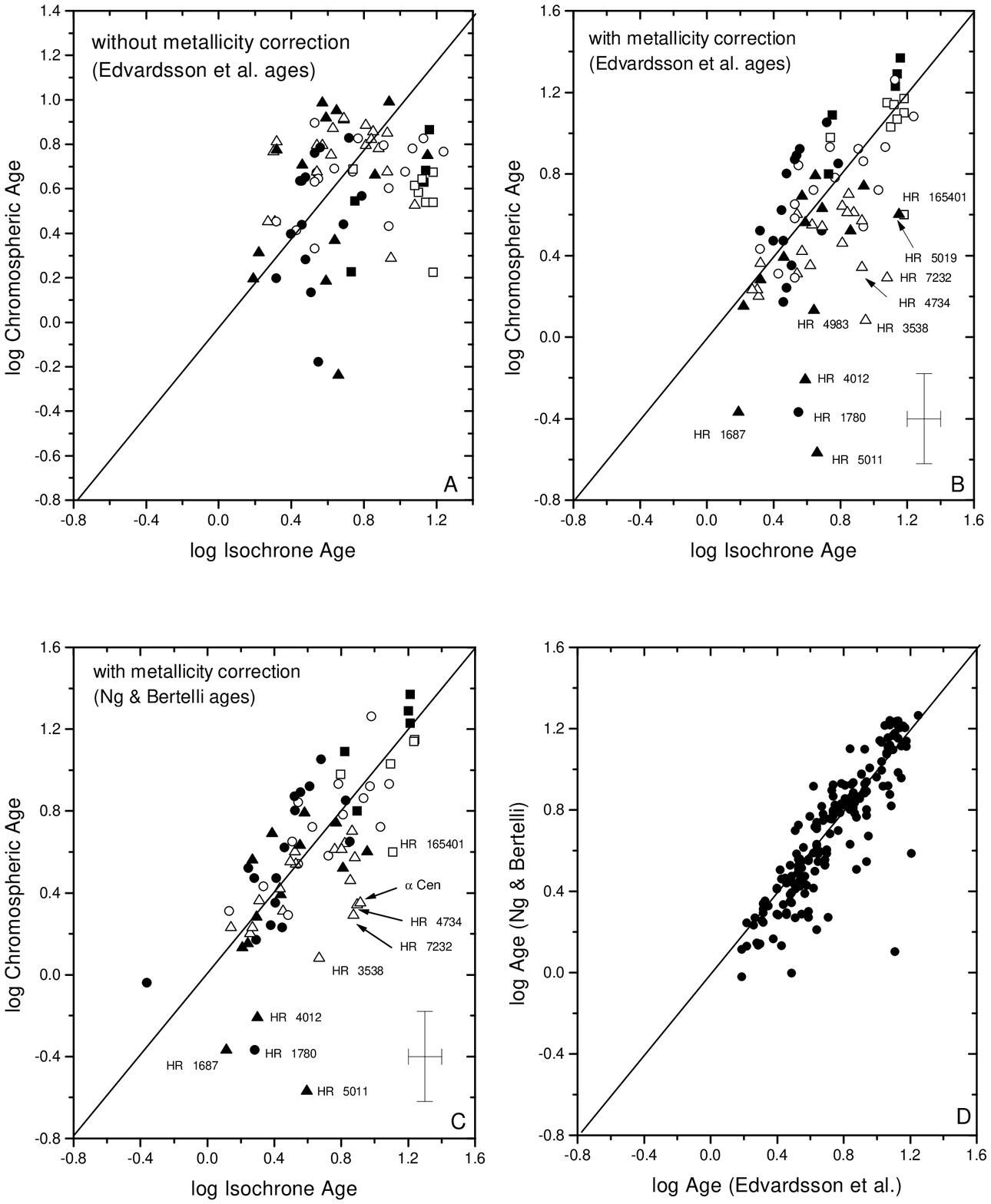}}
      \caption[]{Comparison between chromospheric and isocrone ages for 81 Edv93's 
         stars. The different symbols stand for: stars with [Fe/H] $\le -0.50$ (squares), stars 
         with $-0.50 < {\rm [Fe/H]} \le -0.10$ (circles), and stars with [Fe/H] $>-0.10$ dex. The 
         styles indicate whether the star was already used to correct the chromospheric age scale (open 
         symbols) or not (solid symbols). The upper panels show a comparison between the isochrone and the 
         chromospheric ages, before and after the application of the metallicity corrections. Clearly it can be 
         seen that the metallicity corrections improve substantially the chromospheric ages. The scatter is comparable 
         to the expected error in both methods. Panel c shows the same comparison, now using Ng \& Bertelli 
         (\cite{ng}) ages. Panel d shows a comparison, in the same scale, between the isochrone age determinations 
         by Edv93 and Ng \& Bertelli (\cite{ng}). }
      \label{agecomp}
      \end{figure*}

     Few stars deviated significantly from the expected relation in panels b and c of Figure \ref{agecomp}. 
     With two exceptions (\object{HR 1780} 
     and \object{HD 165401}), all of them have metallicities greater than $-0.05$ dex. The 
     deviation is in the same direction: stars that appear to be very young, according to their 
     chromospheres, are found to be much older from their position in the HR diagram. Three 
     of these stars (\object{HR 4734}, \object{HR 7232} and \object{HD 165401}) were measured only once by 
     the Mount Wilson group, according to Duncan et al. (\cite{duncan}). 
     While their high chomospheric activity could be caused by a periodic active phase 
     in their magnetic cycles, all of the other stars were observed more than a hundred times, over a
     time span of 12 years or more, and 
     errors in the chromospheric indices cannot be invoked to explain their different ages. 

     Some of the deviating stars in panel b are chromospherically young, kinematically old stars 
     (hereafter referred as CYKOS), that have been recently investigated by some of us (Rocha-Pinto, Castilho \& 
     Maciel \cite{RPCM}). Their origin is presently unknown. Poveda et al. (\cite{allen}) have also 
     found objects like these amongst UV Ceti stars, and they proposed they could be low-mass blue stragglers, 
     formed by the coalescence of close binaries. Other explanation could be that these stars are themselves 
     close, unresolved binaries. The high rotation rates of these systems tend to keep the chromospheric 
     activity well beyond the normal age where we would expect it to have been diminished. 
     Moreover, systematic chemical variations between them and the other normal 
     stars could be a cause for some deviations from the relation. However, we have found no such difference, 
     using the abundance ratios provided by Edv93. We can but conclude that these stars do not follow the 
     chromospheric activity--age relation.  
     A possibility to be explored by future investigations is whether the 
     metallicity correction to the chromospheric ages of the metal-rich stars have been overestimated. It is 
     apparent from Figure~1 of Rocha-Pinto \& Maciel (\cite{RPM98}) that some metal-rich stars present 
     great difference between their chromospheric and isochrone ages.

     The existence of anomalies to the chromospheric activity--age calibration does not throw doubt about 
     the use of chromospheric indices to measure stellar 
     ages. The exceptions are few. There are also exceptions to the use of photometric indices to measure 
     stellar parameters, for instance in peculiar stars, but these do not rule out the validity and quality 
     of photometrically calculated parameters. The difference is that there are 
     ways to photometrically distinguish between a normal from a peculiar star. But this is not 
     possible for CYKOS from chromospheric indices alone. We are not sure whether it could be done by 
     photometry. The only way to discover a CYKOS presently is by comparing their activity levels with 
     their isochrone ages and their kinematical characteristics. 

     \subsection{The problem of the metallicity scatter}

\begin{figure*}
      \resizebox{\hsize}{!}{\includegraphics{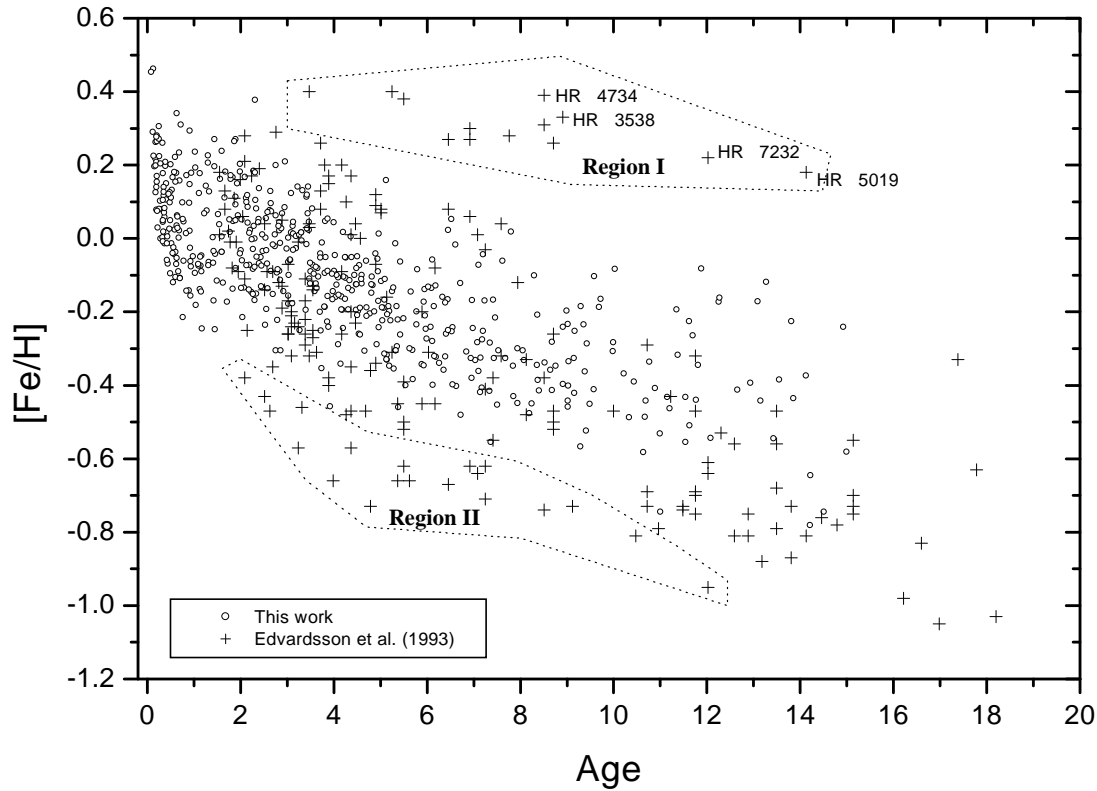}}
      \caption[]{AMRs from Edv93 (crosses) and from this work (open circles). 
        The major differences correspond to two 
        areas in this diagram where we have found no stars, while Edv93 found some. These regions were 
        named I and II, and are marked by dotted lines. The stars labelled have also 
        shown peculiar behaviour in Figure \ref{agecomp}.}
      \label{scatter}
      \end{figure*}

     A more instructive way to compare the metallicity scatter of the two AMR is 
     presented in Figure \ref{scatter}, where we superpose both AMR's. The differences occur mainly in 
     two regions, which we have marked by dotted lines, and named as Region I and II. 
     Region I has also other interesting characteristic: it is well detached from the bulk of 
     stars in both samples. Note that four stars amongst those that have deviated from the mean relation in 
     Figure \ref{agecomp} are also present in Region I. It suggests that the other stars in this location do  
     not follow the chromospheric activity--age relation. We have chromospheric indices for five 
     other stars in this region (\object{HR 3176}, \object{HR 3951}, 
     \object{HR 5423}, \object{HR 8041} and \object{HR 8729}). All of them also 
     present chromospheric ages lower than their isochrone ages, 
     although the age excess is 
     smaller than that found for the stars in Figure \ref{agecomp}. 
     Table \ref{crojoca} presents the logarithmic isochrone age of these stars 
     ($\log I_{\rm age}$), in Gyr, 
     taken from Edv93, and the age excess ($\Delta$), which we define as the 
     logarithmic difference 
     between the isochrone and the chromospheric age of a star (see Rocha-Pinto 
     \& Maciel \cite{RPM98}). For other star, \object{HR 2354}, chromospheric 
     indices were not found in the literature, but emission in the Ca II K line 
     was observed by Hagen \& Stencel (\cite{hagen}). Significant X-ray emission 
     was also observed by the ROSAT observatory (H\"unsch, Schmitt \& Voges 
     \cite{hunsch}), confirming that it is 
     chromospherically active. Its spectral type is G3 III-IV. Although Micela, 
     Maggio \& Vaiana (\cite{micela}) showed that the X-ray activity can be used as 
     an age indicator in giants, no study has ever aimed to calibrate 
     an age relation for them. We have used the relation proposed by Kunte, Rao \& Vahia     
     (\cite{kunte}) to estimate an X-ray age for it. We alert that this relation 
     was calibrated only for main-sequence stars, and its use here is merely 
     illustrative. Using the value for the X-ray luminosity over bolometric 
     luminosity given  
     by (H\"unsch, Schmitt \& Voges \cite{hunsch}), we have an X-ray age of 2.5 Gyr 
     to \object{HR 2354}, which is lower than the isochrone age found by 
     Edv93, just like for the other stars with chromospheric ages. 

     \begin{table}
      \caption[]{Age excesses for stars in Region I and II of Edv93's AMR.}
         \label{crojoca}
         \begin{flushleft}
    {\halign{%
    #\hfil & \qquad\hfil$#$\hfil & \qquad\hfil$#$ \cr
    \noalign{\hrule\medskip}
    Name & \log I_{\rm age} & \Delta \cr
    \noalign{\medskip\hrule\medskip}
         \multispan3 \hfil{\bf Region I}\hfil \cr
         HR 2354$^a$ & 0.84 & 0.44 \cr
         HR 3176 & 0.94  & 0.31 \cr
         HR 3538 & 0.95 &  0.87 \cr
         HR 3951 & 0.81 & 0.48 \cr
         HR 4734 & 0.93 & 0.59 \cr
         HR 5019 & 1.15 & 0.55 \cr
         HR 5423$^b$ & 0.84 & 0.53 \cr
         HR 7232 & 1.08 & 0.79 \cr
         HR 8041$^b$ & 0.89 & 0.48 \cr
         HR 8729 & 0.93 & 0.53 \cr
         \multispan3 \hfil{\bf Region II}\hfil \cr
         HR 4657 & 0.73 & 0.32 \cr
         HR 5447 & 0.32  & -0.17 \cr
         HR 8354 & 0.75  & -0.32 \cr
  \noalign{\medskip\hrule}
  \noalign{$^a$ Age estimated by X-ray luminosity (H\"unsch et al. \cite{hunsch})}
  \noalign{$^b$ Chromospheric indices given by Young et al. (\cite{young})}         }}
         \end{flushleft}
   \end{table}

     The presence of CYKOS in this region 
     also points to a wrong age determination (both isochrone and chromospheric), 
     since as a coalesced star, as an 
     unresolved close binary would have both chromospheric activity and position 
     in the HR diagram uncorrelated 
     with its real age. This can explain why there is a gap between Region I and the other stars. 
     Chromospheric indices for other stars in this region could test more properly this hypothesis. 
     
     The other region marked in Figure \ref{scatter} corresponds to the metal-poor stars with young to intermediate 
     ages. None of such stars are present in our AMR. This could be explained if there 
     were some systematic error in the metallicity corrections to the chromospheric ages that would avoid locating stars 
     in this region. Unfortunately, there are chromospheric measurements for only three stars in Region II. 
     Their age excesses do not follow a systematic trend, as for the stars in Region I, 
     as can be verified in Table \ref{crojoca}. 

     We have checked if the scatter in our AMR could have been partially destroyed due to the use of 
     the metallicity corrections. Rocha-Pinto \& Maciel (\cite{RPM98}) have proposed such a correction 
     from their finding that the difference between chromospheric and isochrone ages were related to 
     the stellar metallicity. The correction itself disregards the scatter in the data. If this scatter is 
     real, probably reflecting different chemical compositions, the use of a calibrated metallicity 
     correction could remove partially the dispersion in the chromospheric AMR. We have used a simulation 
     analogous to that used in the previous section. This time, we have used a `AMR' composed only  
     by scatter, that is, 730 stars were randomly distributed in age (from 0 to 16 Gyr) and metallicity 
     (from $-1.2$ to 0.4 dex). We have taken explicitly into account the dispersion in the metallicity 
     correction, taken from Figure 1 of Rocha-Pinto \& Maciel (\cite{RPM98}), as well as the errors 
     in $\log R'_{\rm HK}$ and [Fe/H]. We have found that the chromospheric AMR preserves closely the real 
     metallicity dispersion of the data, even with the greatest possible scatter 
     for the disk AMR. 

     We have found no indication that the differences in the AMR are due to difference in the methods 
     used to estimate age and metallicity. We are left with the hypothesis that one of the samples (Edv93's  
     or ours) is not suitable to investigate this constraint.

     There are arguments to suspect that Edv93's sample selection has biased their results. 
     Their sample 
     was selected very carefully to allow the study of the abundance ratios evolution. It is not suitable to other 
     studies such as, for instance, the metallicity distribution. Part of the scatter in Edv93's AMR comes from their 
     selection procedure to have nearly 
     equal number of stars in predetermined $\delta m_1$ intervals, to assure a 
     good coverage of [Fe/H] values in their sample. The authors themselves make a correction 
     for this, using a metallicity distribution of 446 dwarfs, after which the 
     metallicity dispersion diminishes to around 0.21 dex. While this value is  
     closer to previous estimates of the metallicity dispersion in the solar 
     vicinity, it is still greater than ours. What is generally not recognized is that the same kind of problem can 
     occur in the investigation of the metallicity dispersion inside each age bin. 

     We have tested this with a 
     number of artificial samples. We have deliberately selected randomly nearly 25 stars in each of the 
     metallicity bins 
     that Edv93 used, amounting 189 stars just like their sample, from a set of 2000 artificial stars. These 
     pre-selected AMRs 
     were compared with other AMRs composed by 189 stars randomly selected from the same parent population. We 
     have verified that contrary to the expectation, {\it this kind of selection procedure does not 
     necessarily imply a metallicity scatter greater than what would be found with a more 
     unbiased selection procedure}. However, there is a significant difference between these simulations and 
     the procedure made by Edv93. According to them, they had observed {\it the 25 brightest stars} in each metallicity 
     box, after disregarding binaries and stars rotating faster than 25 km/s. This probably causes a bias that was 
     not taken into account by the authors. If there was a one-to-one relation between metallicity and age, the age 
     of the 25 brightest stars would be the same as that of the 25 faintest stars. On the other hand, if there is a 
     real metallicity scatter in the interstellar medium, at the same metallicity, the brightest stars would not 
          have the same age dispersion as the faintest stars, or the age dispersion of a subsample randomly selected. 
     The brightest stars will be generally the nearer stars and the most evolved, which are higher in the HR-diagram. 
     From stellar evolutionary theory we also know that these are generally the older stars, or the hotter stars. Since 
     Edv93 also avoid faster  rotating stars (which corresponds to the hotter stars), the majority of these stars will be 
          old.  Therefore, Edv93's metallicity bins have probably non-representative age distributions, which changes 
     the whole dispersion in the AMR.

     This selection effect was introduced by intrinsic observational limitations. Other selection effects are also present 
     in their sample due to this problem. For instance, Garnett \& Kobulnicky (\cite{garnett}) have shown  
     that there is a distance bias in Edv93's sample. Stars at distances 30-80 pc 
     from the Sun are systematically metal-poor. The explanation for this trend is provided by Edv93 themselves. 
     To have equal number of stars in each metallicity bin, they observed the metal-poor stars at a large volume. 
     Garnett \& Kobulnick conclude that the real scatter in the AMR should be lower than 0.15 dex, which 
     favours our results. Edmunds \cite{eddy} favours the same conclusion.  

     However, the puzzle cannot be considered solved. It is highly necessary to have an independent study of the 
     spectroscopic 
     AMR, preferably using a sample as large as that of Twarog. It is also important to make a more 
     detailed investigation of the 
     link between age and stellar activity, especially for the CYKOS and other deviating stars. 
     Presently,  
     we can conclude that the photometric, chromospheric AMR does not give support to the great scatter 
     in the AMR as found by Edv93. Therefore, the AMR can be still regarded as a strong constraint 
     to chemical evolution models, just like the G dwarf metallicity distribution and the radial 
     abundance gradients (see 
     Chiappini, Matteucci \& Gratton \cite{chiappini}).
 
\section{Conclusions}
                   
    We have used an extended sample composed of 552 late-type dwarfs, with chromospheric ages 
    and photometric metallicities, to address 
    the AMR of the solar neighbourhood. Our main conclusions can be summarized 
    as follows:
    
    \begin{enumerate}
    \item The AMR is found to be a smooth function of time. The average 
        metallicity has increased by at least 0.56 dex in the last 12-15 Gyr. No absence of 
        rich stars ageing 3 to 5 Gyr, as suggested by Carraro et al. (\cite{carraro}), 
        was found. The results are in fairly good agreement with the mean points 
        of Edv93's and Ng \& Bertelli (\cite{ng})'s AMR.
    \item The initial metallicity of the disk was around $-0.70$ dex. This suggests  
        some previous enrichment of the gas which gave rise to the thin disk.
    \item A number of very metal rich stars in Edv93 could be composed by chromospherically young, 
        kinematically old stars (CYKOS). These stars could originate from coalesced stars or unresolved 
        close binaries. This may indicate that their isochrone ages, as measured by Edv93, are wrong. This 
        can, at least, explain why Edv93 have found some old stars more metal rich than the Sun.
    \item An average intrinsic metallicity dispersion of 0.13 dex was found. Several hypotheses 
        were tested to explain this scatter compared to that found by Edv93. We have found no 
        indication that it could be produced by our method to estimate ages and metallicities. 
        On the other hand, this dispersion agrees closely with the recent conclusions by 
        Garnett \& Kobulnicky (\cite{garnett}) that the real scatter in the AMR should be 
        lower than 0.15 dex. Nevertheless, additional 
        independent determinations of the AMR are strongly encouraged to confirm this view. 
    \end{enumerate}    
  
    Some important bias in the sample can be investigated with the aid of kinematical data. 
    A third paper on this series addressing this topic is being planned.

\begin{acknowledgements}
      We acklowledge critical reading of the manuscript and suggestions made by Eric Bell. The referee, 
      Dr. David Soderblom has raised several important points, which contributed to 
      improve the paper. This research made extensive use of the SIMBAD database, operated at CDS, 
      Strasbourg, France. We acklowledge support by FAPESP and CNPq to WJM and HJR-P,
      NASA Grant NAG 5-3107 to JMS, and the Finnish Academy to CF. 
\end{acknowledgements}


\begin{thebibliography}{}

   \bibitem[1985]{annkang} Ann H.B., Kang Y.H., 1985, JKAS 18, 79
   \bibitem[1995]{baliunas} Baliunas S.L., Donahue R.A., Soon W.H., et al., 1995, ApJ 438, 269
   \bibitem[1988]{barry} Barry D.C., 1988, ApJ 334, 446

   \bibitem[1989]{basri} Basri G., Wilcots E., Stout N., 1989, PASP 101, 528
   \bibitem[2000]{binney} Binney J.J., Sellwood J.A., 2000, preprint (astro-ph/0003194)
   \bibitem[1985]{carl} Carlberg R.G., Dawson P.C., Hsu T., VandenBerg D.A., 1985, 
       ApJ, 294, 674

   \bibitem[1994]{carchi} Carraro G., Chiosi C., 1994, A\&A 287, 761
   \bibitem[1998]{carraro} Carraro G., Ng Y.K., Portinari L., 1998, MNRAS 296, 1045
   \bibitem[1985]{cayrel85} Cayrel de Strobel G., Bentolila C., Hauck B., 
         Duquennoy A., 1985, A\&AS, 59, 145
   \bibitem[1997]{chiappini} Chiappini C., Matteucci F., Gratton R., 1997,
      ApJ 477, 765
   \bibitem[1997]{clayton} Clayton D.D., 1997, ApJ 484, L67
   \bibitem[1975]{crau} Crawford D.L., 1975, AJ 80, 955
   \bibitem[1992]{cunha} Cunha K., Lambert D.L., 1992, ApJ 399, 586
   \bibitem[1993]{camisa12} de Freitas Pacheco J.A., 1993, ApJ 403, 673
   \bibitem[1993]{don} Donahue R.A., 1993, PhD Thesis, New Mexico State University
   \bibitem[1998]{don98} Donahue R.A., 1998, in Donahue R.A. and Bookbinder J.A., eds., Stellar 
       Systems and the Sun, ASP Conf. Ser. 154, 1235

   \bibitem[1991]{duncan} Duncan D.K., Vaughan A.H., Wilson O.C., et al., 1991, ApJS 76, 383
   \bibitem[1991]{mayor} Duquennoy A., Mayor M., 1991, A\&A 248, 485
   \bibitem[1998]{eddy} Edmunds M.G., 1998, in Friedli D., Edmunds M., Robert G., Drissen L., eds., Abundance 
       Profiles: Diagnostic Tools for Galaxy History, ASP Conf. Series 147, 147
   \bibitem[1993]{Edv} Edvardsson B., Anderson J., Gustafsson B., Lambert D.L., 
      Nissen P.E., Tomkin J., 1993, A\&A 275, 101 (Edv93)
   \bibitem[1993]{friel} Friel E.D., Janes K.A., 1993, A\&A 267, 75
   \bibitem[2000]{garnett} Garnett D.R., Kobulnicky H.A., 2000, ApJ in press (astro-ph/9912031)
   \bibitem[1991]{gimenez} Gim\'enez A., Reglero V., de Castro E., 
      Fern\'andez-Figueroa M.J., 1991, A\&A 248, 563
   \bibitem[1996]{grigsby} Grigsby J.A., Mulliss C.L., Baer G.M., 1996, PASP 108, 953
   \bibitem[1985]{hagen} Hagen W., Stencel R.E., 1985, AJ 90, 120
   \bibitem[1998]{HM98} Hauck B., Mermilliod M., 1998, A\&AS 129, 431
   \bibitem[1996]{HSDB} Henry T.J., Soderblom D.R., Donahue R.A., Baliunas S.L., 1996, 
     AJ 111, 439 (HSDB)
   \bibitem[1975]{HC} Houk N., Cowley A.P., 1975, Michigan Catalog of Two-Dimensional 
     Spectral Types for the HD Stars, Vol. 1, Declinations $-90^o$ to $-53^o$, Ann Arbor, 
     University of Michigan
   \bibitem[1978]{houk78} Houk N., 1978, Michigan Catalog of Two-Dimensional 
     Spectral Types for the HD Stars, Vol. 2, Declinations $-53^o$ to $-40^o$, Ann Arbor, 
     University of Michigan
   \bibitem[1982]{houk82} Houk N., 1982, Michigan Catalog of Two-Dimensional 
     Spectral Types for the HD Stars, Vol. 3, Declinations $-40^o$ to $-26^o$, Ann Arbor, 
     University of Michigan
   \bibitem[1988]{HSM} Houk N., Smith-Moore M., 1988, Michigan Catalog of Two-Dimensional 
     Spectral Types for the HD Stars, Vol. 1, Declinations $-26^o$ to $-12^o$, Ann Arbor, 
     University of Michigan
   \bibitem[1998]{hunsch} H\"unsch M., Schmitt J.H.M.M., Voges W., 1998, A\&AS 127, 251
   \bibitem[1987]{JS} Johnson D.R.H., Soderblom D.R., 1987, AJ 93, 864
   \bibitem[1996]{KG96} Kennicutt R.C., Jr., Garnett, D.R., 1996, ApJ 456, 504
   \bibitem[1996]{KS96} Kobulnicky H.A., Skillman E.D., 1996, ApJ 471, 211
   \bibitem[1988]{kunte} Kunte P.K., Rao A.R., Vahia M.N., 1988, Ap\&SS 143, 207
   \bibitem[1999]{lach} Lachaume R., Dominik C., Lanz T., Habing H.J., 1999, A\&A 348, 89
   \bibitem[1989]{lee} Lee S.-W., Ann H.B., Sung H., 1989, JKAS 22, 43
   \bibitem[1994]{lep94} L\'epine J.R.D., Duvert G., 1994, A\&A 286, 60
   \bibitem[1999]{lep99} L\'epine J.R.D., Sartori M.J., Marinho E.P., 1999, in Stromlo 
      Workshop on High Velocity Clouds, Gibson B.K. and Putman M.E., eds., in press 
   \bibitem[1990]{marsakov} Marsakov V.A., Shevelev Yu.G., Suchkov A.A., 1990, Ap\&SS, 172, 51
   \bibitem[1991]{meu} Meusinger H., Reimann H.-G., Stecklum B., 1991, A\&A, 245, 57
   \bibitem[1998]{meyer} Meyer D.M., Jura M., Cardelli, J.A., 1998, ApJ 493, 222
   \bibitem[1992]{micela} Micela G., Maggio A., Vaiana G.S., 1992, ApJ 388, 171
   \bibitem[1998]{ng} Ng Y.K., Bertelli G., 1998, A\&A 329, 943
   \bibitem[1985]{nissen} Nissen P.E., Edvardsson B., Gustafsson B., 1985, in 
      Production and Distribution of C, N, O Elements, Danziger I.J., Matteucci F. and 
      Kjar K., eds., p. 131
   \bibitem[1984]{noyes} Noyes R.W., Hartmann L.W., Baliunas S.L., Duncan D.K.,
     Vaughan A.H., 1984, ApJ 279, 778

   \bibitem[1983]{olsen83} Olsen E.H., 1983, A\&AS 54, 55
   \bibitem[1984]{olsen84} Olsen E.H., 1984, A\&AS 57, 443
   \bibitem[1988]{olsen88} Olsen E.H., 1988, A\&A 189, 173
   \bibitem[1993]{olsen93} Olsen E.H., 1993, A\&AS 102, 89
   \bibitem[1994]{olsen94} Olsen E.H., 1994, A\&AS 104, 429
   \bibitem[1996]{allen} Poveda A., Allen C., Herrera M.A., Cordero G., Lavalley C., 1996, A\&A 308, 55
   \bibitem[2000]{RPCM} Rocha-Pinto H.J., Castilho B.V., Maciel W.J., 2000, in preparation
   \bibitem[1996]{RPM96} Rocha-Pinto H.J., Maciel W.J., 1996, MNRAS 279, 447
   \bibitem[1998]{RPM98} Rocha-Pinto H.J., Maciel W.J., 1998, MNRAS, 298, 332
   \bibitem[2000]{paperII} Rocha-Pinto H.J., Scalo J., Maciel W.J., Flynn C., 2000, A\&A 
   submitted (Paper II)
   \bibitem[1997]{saar} Saar S.H., Donahue R.A., 1997, ApJ 485, 319
   \bibitem[1989]{schuniss} Schuster W.J., Nissen P.E., 1989, A\&A 221, 65
 
   \bibitem[1985]{soder} Soderblom D.R., 1985, AJ 90, 2103 (S85) 
   \bibitem[1990]{soder90} Soderblom D.R., 1990, AJ 100, 204
   \bibitem[1991]{soder91} Soderblom D.R., Duncan D.K., Johnson D.R.H., 1991, ApJ 375, 
     722
   \bibitem[1998]{SKH} Soderblom D.R., King J.R., Henry T.J., 1998, AJ 116, 396
   \bibitem[1991]{strobel} Strobel A., 1991, A\&A 247, 35

   \bibitem[1980]{twar} Twarog B.A., 1980, ApJ 242, 242
   \bibitem[1985]{VandenBerg} VandenBerg D.A., 1985, ApJS 58, 711
   \bibitem[1977]{wielen} Wielen R., 1977, A\&A 60, 263 
   \bibitem[1996]{WFD} Wielen R., Fuchs B., Dettbarn C., 1996, A\&A 314, 438
   \bibitem[1997]{wiewil} Wielen R., Wilson T.L., 1997, A\&A 326, 139
   \bibitem[1970]{wooley} Wooley R.v.d.R., Epps E.A., Penston M.J., Pocock S.B., 1970, 
      Royal Obs. Annals, No. 5
   \bibitem[1989]{young} Young A., Ajir F., Thurman G., 1989, PASP 101, 1017



\end{thebibliography}
\end{document}